\documentclass[twocolumn,letterpaper,groupedaddress,showpacs,prb]{revtex4}
\usepackage{graphics}
\usepackage{amsmath}
\usepackage{graphicx}
\usepackage{amsfonts}
\usepackage{amssymb}

\DeclareSymbolFont{rsfs}{U}{rsfs}{m}{n}
\DeclareSymbolFontAlphabet{\mathcal}{rsfs}

\begin{document}

\title{Quantum Statistics and Entanglement of Two Electromagnetic 
       Field Modes Coupled via a Mesoscopic SQUID Ring.}
\author{M.J. Everitt}
\author{T.D. Clark}
  \email{t.d.clark@sussex.ac.uk}
\author{P. Stiffell}
\author{H. Prance}
\author{R.J. Prance}
\affiliation{Quantum Circuits Group, 
             School of Engineering, 
             University of Sussex, 
             Brighton,
             Sussex 
             BN1 9QT, U.K.}

\author{A. Vourdas}
\author{J.F. Ralph}
\affiliation{Department of Electrical Engineering and Electronics, 
             Liverpool University,
             Brownlow Hill, 
             LIverpool 
             L69 3GJ, U.K.}

\begin{abstract}
  In this paper we investigate the behaviour of a fully quantum
  mechanical system consisting of a mesoscopic SQUID ring coupled to
  one or two electromagnetic field modes. We show that we can use a
  static magnetic flux threading the SQUID ring to control the
  transfer of energy, the entanglement and the statistical properties
  of the fields coupled to the ring. We also demonstrate that at, and
  around, certain values of static flux the effective coupling between
  the components of the system is large. The position of these regions
  in static flux is dependent on the energy level structure of the
  ring and the relative field mode frequencies, In these regions we
  find that the entanglement of states in the coupled system, and the
  energy transfer between its components, is strong.

\end{abstract}
\maketitle

\section{Introduction}

\label{I}

In  an  earlier  publication~\cite{ClarkDREPPWS98}  we considered  the
interaction of  a quantum  mechanical SQUID  ring (a single  Josephson
weak link, capacitance  $C_{s}$, enclosed  by a thick  superconducting
ring, inductance  $\Lambda_{s}$) with a classical electromagnetic (em)
field.             Using          quasi-classical              Floquet
theory~\cite{Hochstadt1986,shirley65,chu85,ho87,ChuJ91,cohen-tannoudji92:_atom_photon_inter}
to  solve the time dependent  Schr\"{o}dinger  equation (TDSE) for the
SQUID ring, we were able to show that the ring-field interaction could
be very highly non-perturbative in  nature. In essence  this is due to
the ring Hamiltonian~\cite{PranceCPSDR93} containing a cosine
term  (the     Josephson  coupling    energy)   which   can   generate
non-linearities to all orders.   In addition, this Hamiltonian and its
solutions are  $\Phi_{0}\left( =h/2e\right) $-periodic in the external
static magnetic flux $\left(   \Phi_{xstat}\right)  $ applied  to  the
ring. This quantum  non-linearity ensures that energy exchange between
the field and  the  ring is dominated  by  multiphoton absorption (and
emission)       processes~\cite{ClarkDREPPWS98}. As we  have
demonstrated, this is the case even at modest  field amplitudes and at
frequencies   much less than  the  separation  between the ring energy
levels  $\left( \div  h\right) $. In  this  work we showed  that these
energy exchanges  occurred over  very  small regions  in the bias flux
$\Phi_{xstat}$.  The values in $\Phi_{xstat}$ at which these exchanges
take place are  determined by the ring  energy level structure and the
field frequency $\left( \omega_{e}/2\pi\right)   $ and flux  amplitude
$\left(  \Phi_{e}\right)  $. To  be  precise, it  is  in the  exchange
regions that the     energy expectation value  appears   to  jump (for
example, using  a two level  model) between the time-averaged energies
of the ground  and first excited  states of the ring.  Each transition
(exchange)   region  corresponds to the   separation  between the ring
eigenenergies  equalling $n\hbar\omega_{e}$,  $n$ integer, leading  to
multiphoton absorption, or emission, between  the ring and the field.  
It  is in  these   regions that  the   non-linear nature  of  the ring
Hamiltonian becomes manifest  and where strong (and  non-perturbative)
time dependent superpositions occur  between the  original eigenstates
of the ring.

\begin{figure}[tb!]
  \begin{center}
    \resizebox*{0.28\textwidth}{!}{\includegraphics{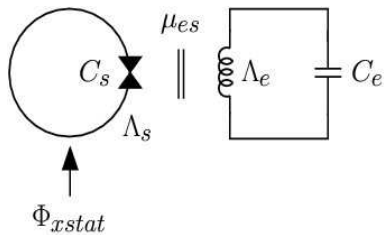}}
  \end{center} 
  \caption{ Block diagram of a SQUID ring coupled to a single em field
mode of frequency $\omega_{e}\left(  =1/\sqrt{C_{e}\Lambda_{e}}\right)  $
where the flux linkage factor, ring to field mode, is $\mu_{es}$. Here, it is
assumed that the temperature $T$ is such that $k_{B}T\ll\hbar\omega_{s}
,\hbar\omega_{e}$ for a SQUID oscillator frequency $\omega_{s}=1/\sqrt{
C_{s}\Lambda_{s}}$ . Also shown is a static bias magnetic flux $\Phi_{xstat}$
applied to the ring.
\label{f1}
}
\end{figure}

Currently there is a great deal  of interest in using mesoscopic SQUID
rings (and other weak   link based circuits) in quantum  technologies,
for                 example,                  in               quantum
computing~\cite{Lopopescuspiller,OrlandoMTvLLM99,MakhlinSS99,AverinNO90}. 
This interest  has been   stimulated by  recent experimental   work on
probing quantum mechanical superposition states in Josephson weak link
circuit
systems~\cite{RouseHL95,SilvestriniRGE00,NakamuraCT97,NakamuraPT99},
and even more  so in the last  year by reports of superposition states
in  SQUID rings~\cite{FriedmanPCTL00,vanderWalWSHOLM00,Cosmelli2001}.  
It seems reasonable to assume that the theoretical description of weak
link systems interacting    with  em fields   (classical  and  quantum
mechanical) is likely to be of great importance  in the development of
any  future superconducting quantum  technologies.  In this regard the
very strong non-linear behaviour exhibited by a single weak link SQUID
ring in  the exchange regions, referred to  above, may prove to  be of
great  utility.  In  order to  test this  viewpoint  we have  recently
considered,   within  a  fully  quantum     mechanical framework,  the
interaction     of a  SQUID      ring   with   an  oscillator    field
mode~\cite{EverittSVVRPP2001}, i.e.  the   simplest coupled system  we
could have chosen (see figure~\ref{f1}).
We found that for the case of the em
field in   a coherent state   the  results derived from  this  quantum
approach compare very  well    with those obtained    previously using
quasi-classical Floquet  theory. In both approaches   the ring and the
field mode only couple  strongly together within the  exchange regions,
i.e over certain narrow regions in  the bias flux $\Phi_{xstat}$. This
means that $\Phi_{xstat}$ can be used  to control the coupling without
losing superposition  coherence in the system. We  note that this work
relates to quantum optical interactions in few  level atoms and to few
level systems involving   either (superconducting) electron  pairs  or
single
electrons~\cite{PranceCPSDR93,SchonZ90,MakhlinSS00,Kastner92,Grabert1992}.

These  initial results for a two  mode (ring + oscillator) system have
encouraged us to draw more parallels with  quantum optics. Rather than
simply consider the SQUID  ring as an electronic  device, we may  also
view it as a tunable, $\Phi_{0}$-periodic, non-linear medium to couple
a  system of quantum oscillators  together.  Regarded as a  non-linear
medium,  there is a clear analogy  to other non-linear quantum systems
in the   context of quantum  optics.  However,  there are  two crucial
differences,  both  of  which may   be of great   importance in future
quantum technologies. First, unlike the SQUID ring, in quantum optical
systems the medium  usually displays a weak  polynomial non-linearity,
even                                   in                       strong
fields~\cite{WodkiewiczE85,YurkeMK86,CamposST89,FearnL89,singer1990,Vourdas92}. 
Second,   all  of  the     properties   of  the  SQUID   (quantum   or
quasi-classical) are $\Phi_{0}$-periodic in bias flux.

\begin{figure}[tb!]
  \begin{center}
    \resizebox*{0.48\textwidth}{!}{\includegraphics{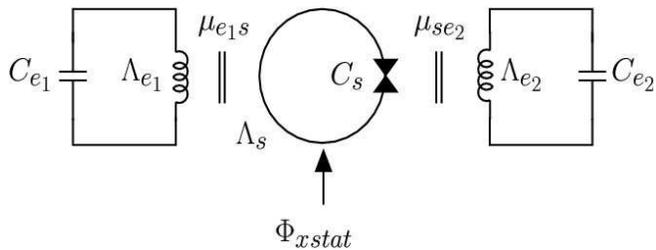}}
  \end{center} 
  \caption{ Block diagram of a SQUID ring coupled to two em field modes of
frequency $\omega_{{e_{1}}}$ and $\omega_{{e_{2}}}$ assuming $k_{B}T\ll
\hbar\omega_{s},\hbar\omega_{{e_{1}}},\hbar\omega_{{e_{2}}}$, with flux
linkage factors $\mu_{e_{1}s}$ and $\mu_{se_{2}}$ between, respectively, the
first field mode and the ring and the ring and the second field mode; all else
as for figure~\ref{f1}.
\label{f2}
}
\end{figure}

In this paper, our objective is to explore the consequences of the
strong quantum non-linearity of the SQUID ring on the interaction, via
the ring, of two oscillator field modes. This arrangement is depicted
in figure~\ref{f2}, 
with the two field modes and the ring oscillator
frequencies taken to be $\omega_{{e_{1}}}/2\pi$,
$\omega_{{e_{2}}}/2\pi$ and $\omega_{s}/2\pi
=\frac{1}{2\pi\sqrt{\Lambda_{s}C_{s}}}$, respectively. As we shall
see, the addition of the second field mode makes this a much more
sophisticated and interesting system than the two mode system
(ring+field oscillator) which was the subject of a recent
publication~\cite{EverittSVVRPP2001}, even though the computational
demands that need to be met are very much greater.  In this regard it
is widely viewed~\cite{Spillsuper,OrlandoMTvLLM99} that the SQUID
ring, as a coherent quantum device, has many potential applications in
the design, development and operation of quantum mechanical circuits
and quantum logic elements. In this paper we consider two aspects of
the quantum behaviour of a SQUID ring which could have a serious
impact in these areas, namely the transfer of entanglement and
frequency conversion between em field modes via the quantum
non-linearity of the ring. In this work we discuss frequency
conversion and entanglement for just two field modes.  However, if the
non-linear aspects of quantum SQUID ring behaviour can be fully
exploited more complicated operations could be envisaged. These may
include using SQUID rings to couple/decouple entangled states in
extended qubit circuit structures and allow frequency conversion
processes between field modes to be modulated, producing coherent
pulse modulated signals. In our opinion, the combination of such
strong non-linear properties, coupled with $\Phi_{0}$-periodic
external bias flux control of this behaviour, makes the SQUID ring
quite unique as a device for application in quantum technologies.
Thus, although the following calculations are concerned with some of
the basic consequences of the quantum interaction of em field modes
with a SQUID ring, we also wish to emphasize the technological
possibilities which may open up as these ring-field mode systems
become more fully understood.

In the work presented here we first consider briefly the two mode
system, including a static bias flux $\Phi_{xstat}$ (figure~\ref{f1}). This
allows us to relate the quasi-classical Floquet approach to the fully
quantum mechanical treatment and demonstrate that our quantum model
can produce consistent results. It also provides the background
formalism for our main goal which is the study of two em field modes
coupled through a SQUID ring. Since our purpose is to study the full
quantum mechanics of the ring-field mode (1 or~2) system, we assume
throughout that the operating temperature $\left( T\right) $ is such
that $\hbar\omega_{{e_{1}},{e_{2}} }\gg k_{B}T$, $\hbar\omega_{s}\gg
k_{B}T$. This ensures that both the ring and field mode(s) behave
quantum mechanically. We then consider the extended quantum circuit,
the two oscillator field modes ($\mathcal{E}_{1}$ and
$\mathcal{E}_{2}$) coupled through a SQUID ring ($\mathcal{S}$) -
figure~\ref{f2} - with a bias flux $\Phi_{xstat}$ also coupled to the ring.
In previous papers~\cite{ClarkDREPPWS98,EverittSVVRPP2001} this bias
flux was used to control the behaviour of the SQUID ring alone or the
ring interacting with one field mode. In the current paper it is used
to control the interaction between two field modes via the non-linear
properties of the SQUID ring. At first sight there might appear to be
no a priori reason why, in this three mode system, it should prove
easy to couple all the components together strongly.  However, at
least for the case of weak inductive coupling between the modes, we
shall show that well characterized energy exchange can take place at
(or close to) certain specific values of $\Phi_{xstat}$. In these
regions of bias flux multiphoton absorption and emission processes
occur.  Thus, the energy required for an interaction to take place is
approximately equal to the energy transfer in the absorption or
emission of an integer number of photons with frequency
$\omega_{{e_{1}}}$ in the first field mode to an integer number of
photons with frequency $\omega_{{e_{2}}}$ in the second.  As for the
two mode (ring+field mode) system, these are the exchange regions
where the effective coupling becomes strong because of the
non-linearity of the SQUID ring. As we shall see, it is in
these regions that many interesting quantum phenomena can be observed.

To emphasize that the coupling across the extended three mode system
is controlled by the bias flux, we calculate the average number of
quanta in each mode and show that there is a large exchange of energy
between the three modes at specific values of $\Phi_{xstat}$, thus
demonstrating that frequency conversion can take place in the system.
In addition, we also calculate the second order correlation
$g_{i}^{(2)}$ (again controlled by the bias flux) which quantifies the
quantum statistics (bunching of quanta) for all three
modes~\cite{ScullyZ97}. We show that as the system evolves in time,
strong entanglement occurs between the three modes. We quantify this
by calculating various entropic quantities based on the von Neumann
entropy~\cite{lindbland1973,lieb_bull1975,wehrl1978,BarnettP91}. These
are chosen for convenience and familiarity and because they can be
used to quantify the degree of entanglement between the subsystems
(field modes and SQUID ring). Although these entropic quantities do
have deficiencies as measures of entanglement~\cite{Vedral1998}, there
is no real consensus about which is the preferred measure within the
quantum technology community. In the absence of any consensus, we opt
for a familiar choice.

\section{The Two Mode Hamiltonian}

\label{II}

\subsection{The SQUID ring in a classical field}

\label{IIA}

\begin{figure}[tb!]
  \begin{center}
    \resizebox*{0.48\textwidth}{!}{\includegraphics{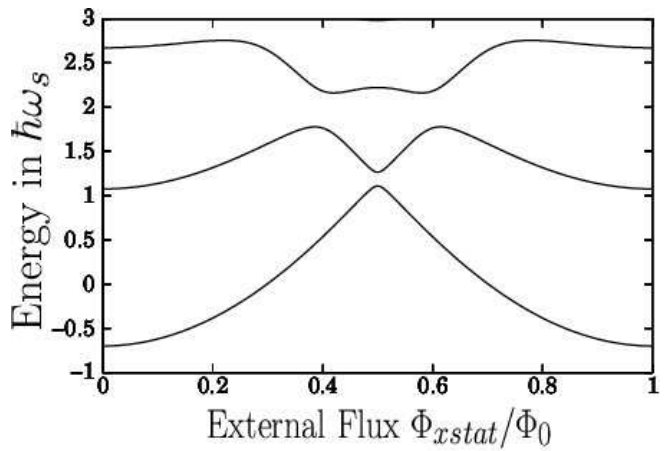}}
  \end{center} 
  \caption{ First three energy eigenvalues $E_{\kappa=0,1,2}$ of a quantum
mechanical SQUID ring as a function of bias flux $\Phi_{xstat}/\Phi_{0}$ over
the range $0\leq\Phi_{xstat}/\Phi_{0}\leq1$ for $C_{s}=1\times10^{-16}F$,
$\Lambda_{s}=3\times10^{-10}H$ $\left(  \hbar\omega_{s}=0.043\Phi_{0}
^{2}/\Lambda_{s}\right)  $ and $\hbar\nu=0.07\Phi_{0}^{2}/\Lambda
_{s}=1.63\hbar\omega_{s}\left(  \nu=1.63\omega_{s}\right)  $.
\label{f3}
}
\end{figure}

\begin{figure}[tb!]
  \begin{center}
    \resizebox*{0.48\textwidth}{!}{\includegraphics{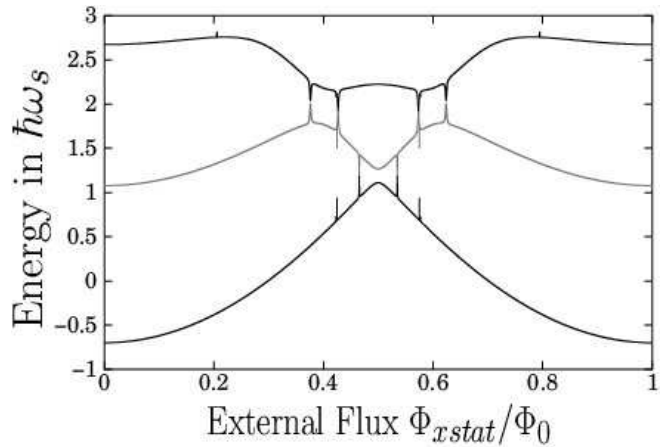}}
  \end{center} 
  \caption{ First three time averaged Floquet energies as a function of
$\Phi_{xstat}/\Phi_{0}$ for the SQUID ring of figure~\ref{f3} where, again,
$\hbar\omega_{s}=0.043\Phi_{0}^{2}/\Lambda_{s}$ $\left(  \frac{\omega_{s}
}{2\pi}\approx1\mathrm{THz}\right)  $ and $\hbar\nu=0.07\Phi_{0}^{2}
/\Lambda_{s}$ $\left(  \nu=1.63\omega_{s}\right)  $ with a classical em field
of frequency$\omega_{xem}=0.5\omega_{s}$ and amplitude $\Phi_{xem}
=2\times10^{-3}\Phi_{0}$ applied. Here, the energy has been normalized to the
ring oscillator energy $\hbar\omega_{s}$.
\label{f4}
}
\end{figure}

In our earlier work we treated the em field
classically~\cite{ClarkDREPPWS98,DigginsWCPPRS97-b} and the SQUID
ring quantum mechanically, using the well known
Hamiltonian~\cite{PranceCPSDR93}
\begin{eqnarray}
H_{s}&=&   \frac{Q_{s}^{2}}
               {2C_{s}}
      +
         \frac{\left(  \Phi_{s}-\left[  \Phi_{xstat}+\Phi_{xem}\sin\omega_{e}t\right]  \right)  ^{2}}
                                              {2\Lambda_{s}}
      \nonumber \\ && -
          \hbar\nu\cos\left(  2\pi\frac{\Phi_{s}}{\Phi_{0}}\right)  \label{eq:HamTDSE}
\end{eqnarray}
Here, $\Phi_{s}$, the magnetic flux threading the ring, and $Q_{s}$,
the electric displacement flux between the electrodes of the weak link
in the ring, are the conjugate variables~\cite{widom79,WidomC82-c} for
the system (with $\left[ \Phi_{s},Q_{s}\right] =i\hbar$), $\hbar\nu/2$
is the matrix element for Josephson pair tunnelling through the weak
link and $\Phi_{xem}$ is the amplitude of the classical magnetic flux
at the ring due to the em field mode. With $\Phi_{xem}$ set to zero we
can solve the time independent Schr\"{o}dinger equation to find the
eigenvalues of the SQUID ring alone as a function of applied flux
$\Phi_{xstat}/\Phi_{0}$. As an example, we show in figure~\ref{f3} 
the first three eigenenergies of the ring [$E_{\kappa=0,1,2}\left(
  \Phi_{xstat}/\Phi_{0}\right) $, where $\kappa =0$ denotes the ground
state, etc.] over the range $0\leq \Phi_{xstat}/\Phi_{0}\leq1$ using
parameters typical of a quantum regime SQUID
ring~\cite{ClarkDREPPWS98,EverittSVVRPP2001}, i.e. $C_{s}
=1\times10^{-16}$F, $\Lambda_{s}=3\times10^{-10}$H (hence $\hbar\omega
_{s}=0.043\Phi_{0}^{2}/\Lambda_{s}$ or
$\frac{\omega_{s}}{2\pi}=9.8\times 10^{11}\mathrm{Hz}$) and
$\hbar\nu=0.07\Phi_{0}^{2}/\Lambda_{s}=1.63\hbar \omega_{s}\left(
  \nu=1.63\omega_{s}\right) $. With $\Phi_{xem}$ turned on, we can use
(\ref{eq:HamTDSE}) to solve the corresponding TDSE. Again, by way of
illustration, we show in figure~\ref{f4} 
the computed time averaged ring
energy expectation values for the first three Floquet states
(eigenvalues of the evolution operator after one period of em field
evolution) as a function of $\Phi_{xstat}/\Phi_{o}$ using the ring
parameters of figure~\ref{f3}. Here, $\omega_{e}$ has been set at
$0.5\omega_{s}$ with the associated $\Phi
_{xem}=2\times10^{-3}\Phi_{0}$. As can be seen, energy exchange
between these time averaged energies occurs at specific values of the
bias flux $\Phi_{xstat}$ and, as we have already pointed out, the
number and position in $\Phi_{xstat}$ of these exchange regions
depends on $\omega_{e}$, $\Phi_{xem}$ and the energy level structure
of the ring. We have observed that these transition (exchange) points
occur for values of bias flux such that (at least for small em field
amplitudes $\Phi_{xem}$) $N\hbar\omega_{e}\approx
E_{i}(\Phi_{x})-E_{j}(\Phi_{x})$ where
$N=0,\pm1,\pm2,\ldots$~\cite{ClarkDREPPWS98}. We note that for the
SQUID ring we can write down a renormalized oscillator frequency
$\Omega_{s}=\omega_{s}+4\hbar^{2}
\pi^{2}\nu\Phi_{0}^{-2}C_{s}^{-1}\omega_{s}^{-1}$ which is related to
the fact that there is an $a_{s}^{\dagger}a_{s}$ term in a Taylor
expansion of the cosine (Josephson) term in the ring
Hamiltonian(\ref{eq:HamTDSE}).

\subsection{The SQUID ring in a non-classical field\label{sec:ncSQUID}
\label{IIB}}

In the fully quantum description the Hamiltonian $H_{t}$ for the SQUID ring-em
oscillator mode system can be written as~\cite{EverittSVVRPP2001}
\begin{equation}
H_{t}=H_{e}+H_{s}-H_{es}. \label{eq:2total}
\end{equation}
where $H_{e}$ and $H_{s}$ are, respectively, the Hamiltonian contributions for
the field and the ring and $H_{es}$ is the interaction energy linking these together.

Following equation (\ref{eq:HamTDSE}), the Hamiltonian for the SQUID
ring alone is~\cite{PranceCPSDR93}
\begin{equation}
H_{s}=\frac{Q_{s}^{2}}{2C_{s}}+\frac{\left(  \Phi_{s}-\Phi_{xstat}\right)
^{2}}{2\Lambda_{s}}-\hbar\nu\cos\left(  2\pi\frac{\Phi_{s}}{\Phi_{0}}\right)
. \label{eq:HamS}
\end{equation}

while the Hamiltonian for the em field (modelled as a parallel capacitance
$\left(  C_{e}\right)  $ - inductance $\left(  \Lambda_{e}\right)  $ cavity
mode equivalent circuit with infinite parallel resistance on resonance) takes
the form $H_{e}=\frac{Q_{e}^{2}}{2C_{e}}+\frac{\Phi_{e}^{2}}{2\Lambda_{e}}$.
Here, $\Phi_{e}$ and $Q_{e}$ are, respectively, the cavity mode magnetic flux
and charge operators for a field mode frequency $\omega_{e}=1/\sqrt
{C_{e}\Lambda_{e}}$. This cavity mode is coupled inductively to the SQUID ring
with a coupling energy $H_{es}=\frac{\mu_{es}}{\Lambda_{s}}\left(  \Phi
_{s}-\Phi_{xstat}\right)  \Phi_{e}$, where $\mu_{es}$ is the em field-SQUID
ring flux linkage factor.

By making a suitable transformation [using the unitary translation operator
${\mathbb{T}=\exp\left(  -i\Phi_{xstat}Q_{s}/\hbar\right)  }$] the Hamiltonian
(\ref{eq:HamS}) - now in calligraphic script - can be written more
conveniently as
\begin{equation}
\mathcal{H}_{s}={\mathbb{T}^{\dagger}H_{s}\mathbb{T}=\frac{Q_{s}^{2}}{2C_{s}
}+\frac{\Phi_{s}^{2}}{2\Lambda_{s}}-\hbar\nu\cos\left(  2\pi\frac{\Phi
_{s}+\Phi_{xstat}}{\Phi_{0}}\right)  } \label{eq:HamST}
\end{equation}
while the Hamiltonian for the em field mode remains unaffected. However, the
interaction energy does transform to $\mathcal{H}_{es}=\frac{\mu_{es}}
{\Lambda_{s}}\Phi_{s}\Phi_{e}$. We denote the magnetic flux dependent
eigenstates of $\mathcal{H}_{s}$ by $|\sigma\rangle_{\mathcal{S}}$. In our
computations we then use a truncated energy eigenbasis both for the ring
($\left|  \sigma\right\rangle $)and the em field mode ($\left|  n\right\rangle
$). The basis states $\left|  \sigma\right\rangle ,\left|  n\right\rangle $,
where $\sigma=\alpha,\ldots,\Omega$ and $n=0,\ldots,N$, are taken so that
$\Omega$ and $N$ are, respectively, much greater than the average number of
quanta in the ring and em field.

\begin{figure}[tb!]
  \begin{center}
    \resizebox*{0.48\textwidth}{!}{\includegraphics{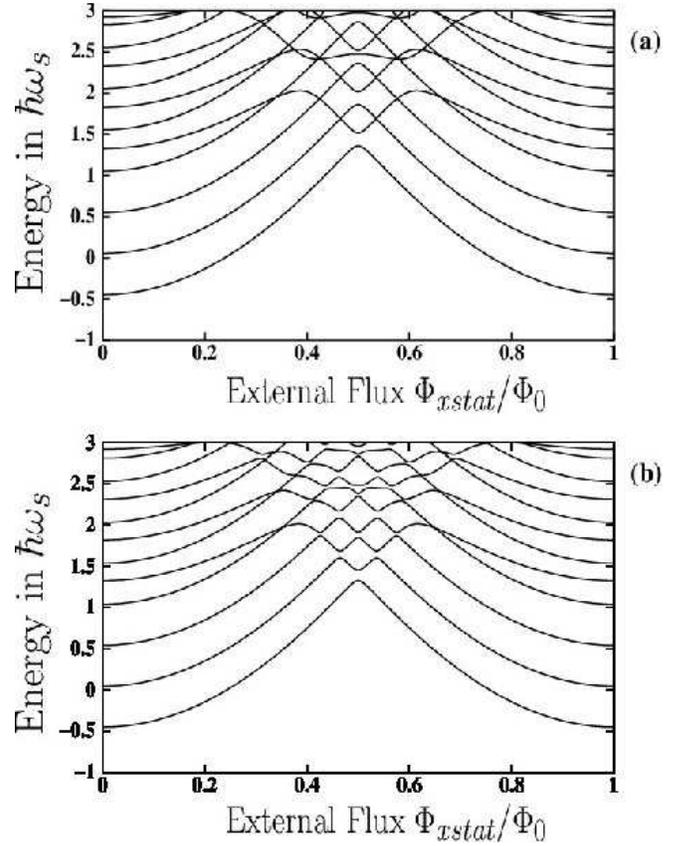}}
  \end{center} 
  \caption{Eigenenergies of the SQUID ring-em field mode system
(normalized in units of $\hbar\omega_{s}$) versus $\Phi_{xstat}/\Phi_{0}$
using the ring parameters of figure~\ref{f3} with $C_{s}=C_{e}=1\times10^{-16}F$,
$\omega_{s}=2\omega_{e}$ $\left(  \mathrm{where}\text{ }\frac{\omega_{s}}
{2\pi}\approx1THz\right)  $ and $\nu=1.63\omega_{s}$. In (a) the field mode-
SQUID ring flux linkage factor $\mu_{es}=0$; in (b) $\mu_{es}=0.1$.
\label{f5}
}
\end{figure}

Using this truncated basis, we can then solve

\begin{equation}
\mathcal{H}_{t}\left|  \xi_{n}\right\rangle =\Xi_{n}\left|  \xi_{n}
\right\rangle \label{eq:eigen}
\end{equation}
to obtain the eigenfunctions and eigenenergies of the two mode system
Hamiltonian $\mathcal{H}_{t}$. The eigenenergies for the ring-field mode
system are shown in figure~\ref{f5}
for the $C_{s}$, $\Lambda_{s}$ and $\hbar\nu$
values used for figure~\ref{f3}, with $\omega_{s}=2\omega_{e}$, as in figure~\ref{f4}. Here,
in figure~\ref{f5}(a) the field mode-ring linkage factor $\mu_{es}
=0.0$ while in figure~\ref{f5}(b) it is $0.1$. From this spectral
decomposition we can form the evolution operator via

\begin{equation}
U\left(  t\right)  =\sum_{n}\left|  \xi_{n}\right\rangle \exp\left(
-\frac{i\Xi_{n}t}{\hbar}\right)  \left\langle \xi_{n}\right|
\label{eq:evolution}
\end{equation}
The time averaged energy expectation values $\langle\langle\mathcal{H}
_{i=s,e}\rangle\rangle$ for the ring~\cite{ClarkDREPPWS98} and the
field (i.e. $\mathcal{H}{{_{s}}}$ and $\mathcal{H}_{{e}}$) can then be
calculated using the expression

\begin{equation}
\langle\langle\mathcal{H}_{i}\rangle\rangle=\lim_{\tau\rightarrow\infty
}\frac{1}{\tau}\int_{0}^{\tau}\mathrm{Tr}[\rho_{i}(t)\mathcal{H}_{i}]dt
\label{eq:average}
\end{equation}
where $\rho_{i},$ $i=e,s$ are the reduced density operators for the em mode
and the SQUID ring, respectively. In practice, we have been able to ensure the
convergence of the integral (\ref{eq:average}) by integrating numerically from
$0$ up to $20,000/\omega_{s}$.

\begin{figure}[tb!] 
  \begin{center} 
    \resizebox*{0.48\textwidth}{!}{\includegraphics{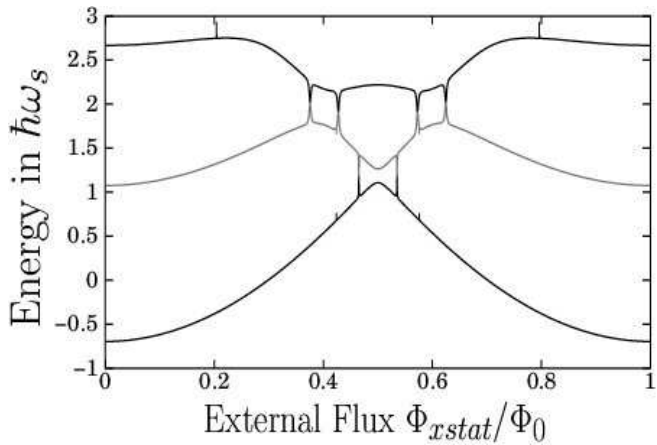}}
  \end{center} 
  \caption{Time averaged energy expectation values of $\mathcal{H}_{s}$
(in units of $\hbar\omega_{s}$) as a function of $\Phi_{xstat}/\Phi_{0}$ for
comparison with figure~\ref{f4} where the em-field mode is in a coherent state and
the SQUID ring is in one of its first three energy eigenstates, i.e. the
initial states are $|\alpha=i\sqrt{20}\rangle_{\mathcal{E}}\otimes
|\sigma\rangle_{\mathcal{S}},$ $\left(  \sigma=0,1,2\right)  $. Here, as for
figure~\ref{f5}, $C_{s}=C_{e}=1\times10^{-16}F$, $\omega_{e}=0.5\omega_{s}\left(
\mathrm{where}\text{ }\frac{\omega_{s}}{2\pi}\approx1\mathrm{THz}\right)  $,
$\nu=1.63\omega_{s}$ but with $\mu_{es}=7.6\times10^{-4}$ for comparison with
figure~\ref{f4}.
\label{f6}
}
\end{figure}

\begin{figure*}[tb!]
  \begin{center}
    \resizebox*{0.9\textwidth}{!}{\includegraphics{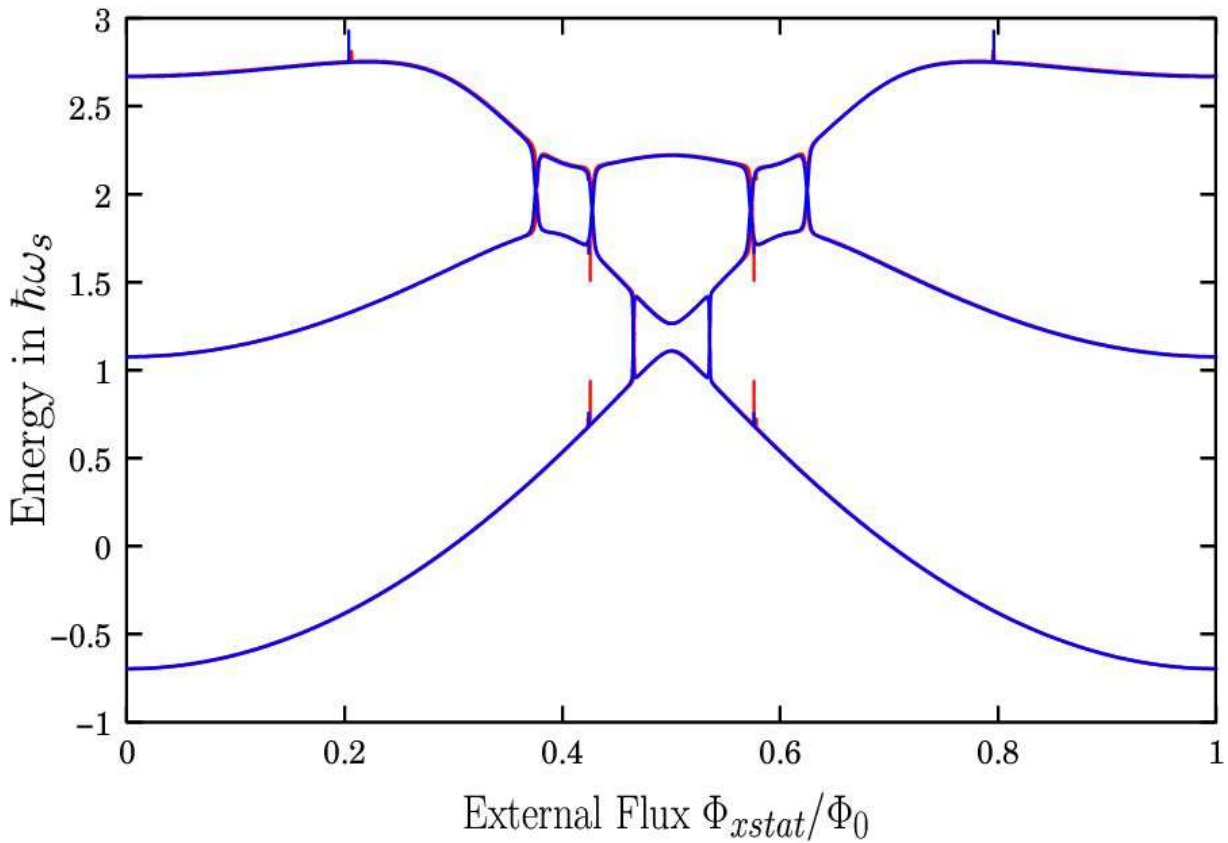}}
  \end{center} 
  \caption{ Comparison between time averaged energies of figures~\ref{f4} (red)
and~\ref{f6} (blue).
\label{f7}
}
\end{figure*}

Clearly, provided we select the correct initial state for the em field 
in the two mode system, there should exist a correspondence between
the Floquet method used in section \ref{IIA} [using
(\ref{eq:HamTDSE})] and the result of a fully quantum mechanical
calculation. As an initial state the coherent state is an obvious
choice since it is the closest quantum state to a monochromatic em
field, as used in the Floquet approach above. With this choice we 
would expect a reasonable agreement between the fully quantum and
quasi-classical computations. In fact the match between  these two
approaches can be very good.  To compare with the quasi-classical
result of section \ref{IIA}, we set the zero time $\left( t=0\right) $
product state as $|\alpha=i\sqrt{20}
\rangle_{\mathcal{E}}\otimes|\sigma\rangle_{\mathcal{S}}$, where
$|\alpha\rangle_{\mathcal{E}}$ is a coherent state of the field
($a_{e}
|\alpha\rangle_{\mathcal{E}}=\alpha|\alpha\rangle_{\mathcal{E}}$ ).
Using this coherent state we show in figure~\ref{f6} 
the calculated $\left\langle \left\langle
    \mathcal{H}_{{s}}\right\rangle \right\rangle $ for an integration
time $\tau=2\times10^{4}/\omega_{s}$ with the energies normalized in
units of $\hbar\omega_{s}$. The computations have been made over the
range $0\leq \Phi_{xstat}/\Phi_{0}\leq1$ for the values of
$\sigma=0,1,2$, using the SQUID ring capacitance, inductance and
$\Phi_{x}$-dependent energy level structure of figure~\ref{f4} . Here, as for
figure~\ref{f3}, we have made $\omega_{e}=0.5\omega_{s}$ while setting the
flux linkage factor $\mu_{es}=0.00076$. We chose this value of
$\mu_{es}$ so that the amplitude of oscillation of the coherent state
in the em field coupled to the SQUID ring is equivalent to that used
in the Floquet calculation of figure~\ref{f4}. It is apparent that the time
averaged energy expectation values, and their exchange regions,
calculated in the quantum model as a function of $\Phi_{xstat}$ are
very close to those found using the quasi-classical Floquet approach.
To emphasise this, we show in figure~\ref{f7}
 the two calculations
superimposed. The very close fit between the two gives us confidence
that the quantum model is physically valid with an accurate
correspondence to the quasi-classical regime when a coherent state is
chosen for the em field. 
 
As we have already noted for the SQUID ring in a classical field (section
\ref{IIA}), we observed that transition (or strong coupling) regions occur
when $N\hbar\omega_{e}\approx E_{i}(\Phi_{x})-E_{j}(\Phi_{x}).$ From figure~\ref{f5}
we can see that this situation exists when degeneracies in the spectrum of the
system are lifted due to the coupling term between the Hamiltonians for the
SQUID ring and em field.

\section{The Three Mode Hamiltonian}

\label{III}

Following on from (\ref{eq:2total}), we now consider a SQUID ring with
oscillator frequency $\omega_{s}$, threaded, as before, by a static bias flux
$\Phi_{xstat}$, but now coupled to two em field modes of frequency
$\omega_{{\ e_{1}}}$ and $\omega_{{e_{2}}}$ (see figure~\ref{f2}). With the usual
flux $\left(  \Phi_{i}\right)  $-charge $\left(  Q_{i}\right)  $ commutation
relation $\left[  \Phi_{i},Q_{j}\right]  =i\hbar\delta_{ij}$, the total
Hamiltonian $\mathcal{H}_{\mathcal{T}}$ for this system can be written as

\begin{equation}
\mathcal{H}_{\mathcal{T}}=\mathcal{H}_{e_{1}}+\mathcal{H}_{e_{1}s}
+\mathcal{H}_{s}+\mathcal{H}_{se_{2}}+\mathcal{H}_{e_{2}}. \label{eq:3total}
\end{equation}
where the SQUID ring Hamiltonian is given by (\ref{eq:HamST}) which has been
transformed into the $\Phi_{xstat}$ basis. We choose to write our component
Hamiltonians in (\ref{eq:3total}) in terms of the annihilation $\left[
a_{i}=\frac{1}{\sqrt{2}}\left(  x_{i}+ip_{i}\right)  \right]  $ and creation
$\left[  a_{i}^{\dag}=\frac{1}{\sqrt{2}}\left(  x_{i}-ip_{i}\right)  \right]
$ operators as
\begin{eqnarray}
\mathcal{H}_{e_{1}}  &  =\hbar\omega_{{e_{1}}}\left(  a_{{e_{1}}}^{\dag
}a_{{\ e_{1}}}+\frac{1}{2}\right) \nonumber\\
\mathcal{H}_{e_{2}}  &  =\hbar\omega_{{e_{2}}}\left(  a_{{e_{2}}}^{\dag
}a_{{\ e_{2}}}+\frac{1}{2}\right) \nonumber\\
\mathcal{H}_{s}  &  =\hbar\omega_{s}\left(  a_{s}^{\dag}a_{s}+\frac{1}
{2}\right)  -\nonumber\\
&  \hbar\nu\cos\left(  \frac{2\pi}{\Phi_{o}}\sqrt{\frac{\hbar}{2C_{s}
\omega_{s}}}\left(  a_{s}^{\dag}+a_{s}\right)  +2\pi\frac{\Phi_{x}}{\Phi_{0}
}\right) \nonumber
\end{eqnarray}
where the position and momentum operators can be defined in terms of the
magnetic flux and the charge operators via $x_{i}=\sqrt{C_{i}\omega_{i}/\hbar
}\Phi_{i}$ and $p_{i}=\sqrt{1/C_{i}\hbar\omega_{i}}Q_{i}$ for oscillator
frequencies $\omega_{i}=1/\sqrt{C_{i}\Lambda_{i}}$, with the subscript $i$
denoting ${e_{1}},{e_{2}}$ for the fields$\mathrm{\ }$or $s$ for the
ring. Hence the Hamiltonians of the components of the system are identical
(but extended to include an extra field mode) to those used in \ref{IIB}, but
written in terms of creation and annihilation operators.

The interaction energies $\mathcal{H}_{e_{1}s}$ and $\mathcal{H}_{se_{2}}$ in
(\ref{eq:3total}), each of which represents the inductive coupling between the
SQUID ring and the oscillator modes ${e_{1}}$ and ${e_{2}}$, respectively, are
given by
\begin{eqnarray}
\mathcal{H}_{e_{1}s}  &  =-\hbar\omega_{s}\frac{\mu_{e_{1}s}}{2}\sqrt{\frac{
C_{s}\omega_{s}}{C_{{e_{1}}}\omega_{{e_{1}}}}}\left(  a_{s}^{\dag}
+a_{s}\right)  \left(  a_{{e_{1}}}^{\dag}+a_{{e_{1}}}\right)
\nonumber\label{eq:interactions}\\
\mathcal{H}_{se_{2}}  &  =-\hbar\omega_{s}\frac{\mu_{se_{2}}}{2}\sqrt{\frac{
C_{s}\omega_{s}}{C_{{e_{2}}}\omega_{{e_{2}}}}}\left(  a_{s}^{\dag}
+a_{s}\right)  \left(  a_{{e_{2}}}^{\dag}+a_{{e_{2}}}\right) \nonumber
\end{eqnarray}

\section{Time Evolution of the Three Mode System}

\label{IV}

The Hilbert space $\mathcal{T}$ for the SQUID ring-em field system is a tensor
product of the Hilbert space $\mathcal{S}$ for the ring and the Hilbert spaces
$\mathcal{E}_{1}$ and $\mathcal{E}_{2}$ for the fields, that is to say
$\mathcal{T}=\mathcal{E}_{1}\otimes\mathcal{S}\otimes\mathcal{E}_{2}$. We
denote in roman script the simple harmonic, em oscillator mode number
eigenstates $|n\rangle_{i}$ ($a_{i}^{\dagger}a_{i}|n\rangle_{i}=n|n\rangle
_{i}$). In representing the SQUID ring we use greek script, for example
$|\alpha\rangle_{s},|\beta\rangle_{s},|\gamma\rangle_{s},\ldots,$ to represent
the eigenstates of the Hamiltonian $\mathcal{H}_{s}$ in order to distinguish
these from the number eigenstates $|n\rangle_{s}$ of the ring $\left(
a_{s}^{\dagger}a_{s}|n\rangle_{s}=n|n\rangle_{s}\right)  $.

In dealing with the time evolution of the coupled three mode system we must
first solve the eigenproblem [see (\ref{eq:eigen})] . As in section \ref{II}
for the two mode system, we use a truncated basis. This has the form

\begin{align}
\big\{&  \left|  N_{n,\kappa,m}\right\rangle    \equiv 
        \left|  n\right\rangle_{{\mathcal{E}}_{1}}\otimes\left|  \kappa\right\rangle _{\mathcal{S}}
\otimes\left|  m\right\rangle _{\mathcal{E}_{2}}|
\nonumber\\
  &n    =0,...,N_{1},\,\kappa=\alpha,...,\Omega,\,m=0,...,N_{2}\big\}  .
\label{eq:basis}
\end{align}
where $\left|  n\right\rangle _{{\mathcal{E}}_{1}}$ and $\left|
m\right\rangle _{\mathcal{E}_{2}}$ are the number states for the two field
modes and $\left|  \kappa\right\rangle _{\mathcal{S}}$ are the energy
eigenstates of the SQUID ring Hamiltonian. Here, $N_{1}$, $\Omega$ and $N_{2}$
are taken to be much greater than the average number of quanta in each
component of the system. With the eigenfunctions and eigenenergies of
(\ref{eq:eigen}), but using the three mode Hamiltonian, the evolution operator
can be calculated using the expression (\ref{eq:evolution}). Then, assuming
that the system at $t=0$ is described by the density matrix $\rho(0)$, the
density matrix $\rho(t)=U(t)\rho(0)U^{\dagger}(t)$ at a later time $t$ can be
found, as can the reduced density matrices $\rho_{e_{1}}=\mathrm{Tr}
_{{\mathcal{S}\otimes{\mathcal{E}_{2}}}}\left(  \rho\right)  $, $\rho
_{s}=\mathrm{Tr}_{{{\mathcal{E}}_{1}}\otimes{\mathcal{E}_{2}}}\left(
\rho\right)  $, $\rho_{{e_{2}}}=\mathrm{Tr}_{{\mathcal{E}_{1}}\otimes
\mathcal{S}}\left(  \rho\right)  $ and $\rho_{{e_{1}}s}=\mathrm{Tr}
_{{\mathcal{E}_{2}}}\left(  \rho\right)  $, $\rho_{s{e_{2}}}=\mathrm{Tr}
_{{\mathcal{E}_{1}}}\left(  \rho\right)  $, $\rho_{{e_{1}}{e_{2}}}
=\mathrm{Tr}_{\mathcal{S}}\left(  \rho\right)  $. With these density matrices
determined, we can then investigate a range of parameters which reveal much of
the quantum behaviour of the three mode system, for example, the von Neumann entropy.

In the following sections of the paper we present numerical results
demonstrating various aspects of this behaviour. Since the examples given are
intended to be illustrative in nature, for simplicity we have made the
capacitances for all three modes of the system the same, these being typical
of quantum regime oscillators operating at a few K, i.e. $C_{{e_{1}}
}=C_{{\ e_{2}}}=C_{s}=10^{-16}\mathrm{F}$. Amongst other things we wish to
show that quantum frequency conversion can occur between the two oscillator
field modes, via the SQUID ring. We have therefore made the two mode
frequencies differ by a factor of two, i.e. $\omega_{{e_{1}}}=2\omega_{{e_{2}
}}$ while again, for simplicity, setting $\omega_{{e_{1}}}=\omega_{s}$. With
all capacitances identical these frequencies correspond to $\Lambda_{{e_{1}}
}=\frac{1}{4}\Lambda_{{e_{2}}}=\Lambda_{s}$ for a typical SQUID ring
inductance $\Lambda_{s}=3\times10^{-10}\mathrm{H}$ (see section \ref{II}).
Again, as in section \ref{II}, we have put $\hbar\nu=0.07\Phi_{0}^{2}
/\Lambda_{s}\,\left(  \nu=1.63\omega_{s}\right)  $ which is typical value of
the pair tunnelling matrix element for quantum regime SQUID rings operating at
a few K. In addition, we have set the ring-field mode flux linkage factors at
$\mu_{{e_{1}}s}=0.01$ and $\mu_{s{e_{2}}}=0.1$\ which approximates to some
reported experiments in the literature involving two oscillator field modes
coupled through a SQUID ring~\cite{WhitemanSCPPDR98,WhitemanCPPSRD98}.

\subsection{Strong Coupling of the SQUID ring to em field modes}

\label{IVA}

\begin{figure}[tb!]
  \begin{center}
    \resizebox*{0.48\textwidth}{!}{\includegraphics{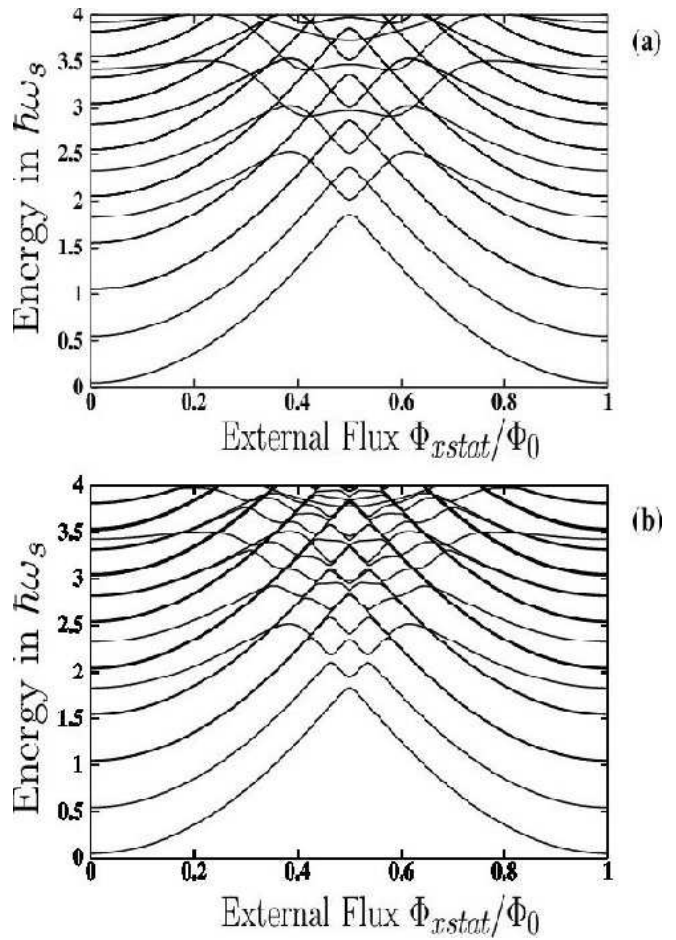}}
  \end{center} 
  \caption{ Eigenenergies (in units of $\hbar\omega_{s}$) versus
$\Phi_{xstat}/\Phi_{0}$ of the three mode (em mode-ring-em mode) system with
$C_{s}=C_{{e_{1}}}=C_{{e_{2}}}=1\times10^{-16}F$, $\omega_{s}=\omega
_{{\ e_{1}}}=2\omega_{{e_{2}}}\left(  \mathrm{where}\text{ }\frac{\omega_{s}
}{2\pi}\approx1\mathrm{THz}\right)  $, $\nu=1.63\omega_{s}$ and flux linkage
factors $\mu_{e_{1}s}$ and $\mu_{se_{2}}$. In (a) $\mu_{e_{1}s}$ and
$\mu_{se_{2}}$ are set equal to zero while in (b) $\mu_{e_{1}s}=0.01$ and
$\mu_{se_{2}}=0.1$.
\label{f8}
}
\end{figure}

In section \ref{II} we demonstrated that strong coupling between the SQUID
ring and the em field occurs when degeneracies in the spectrum of the
Hamiltonian are lifted due to the coupling between the components of the
system. Numerically, this equates to the condition $N\hbar\omega_{e}\approx
E_{i}(\Phi_{x})-E_{j}(\Phi_{x})$, where $N$ is integer and the $E_{i}(\Phi
_{x})$ and $E_{j}(\Phi_{x})$ are the $i^{th}$ and $j^{th}$ eigenvalues of the
system Hamiltonian $\mathcal{H}_{t}$. These regions of strong coupling (the
exchange regions), which develop at specific values of the bias flux $\left(
\Phi_{xstat}\right)  $ applied to the SQUID ring, are dependent on the ring
eigenenergy structure and the frequency of the em field. Similarly, for the
three mode system strong energy exchange will occur between the two field
modes and the SQUID ring when the coupling terms lift degeneracies in the
spectrum of the Hamiltonian. This will occur when
\begin{equation}
N_{{e_{1}}}\hbar\omega_{{e_{1}}}-N_{{e_{2}}}\hbar\omega_{{e_{2}}}\approx
E_{i}(\Phi_{x})-E_{j}(\Phi_{x}) \label{eq:match}
\end{equation}
where, now, the $E_{i,j}(\Phi_{x})$ are the $i^{th}/j^{th}$ eigenvalues of
$\mathcal{H}_{\mathcal{S}}$ and $N_{{e_{1}},{e_{2}}}$ are integers (positive
or negative). In this case $N_{{e_{1}}}$ photons with frequency $\omega
_{{e_{1}}}$ in the first field mode are used to excite the SQUID from the
$j^{th}$-state to the $i^{th}$-state with the emission of $N_{{e_{2}}}$
photons of frequency $\omega_{{e_{2}}}$ into the second field mode. Taking the
ring-field parameters set above, we have calculated the energy eigenvalues for
the three mode system. These are shown in figure~\ref{f8}.
 It can be seen that these
eigenenergies possess a very rich structure which leads directly to the
results presented in this paper.

\begin{figure}[tb!]
  \begin{center}
    \resizebox*{0.48\textwidth}{!}{\includegraphics{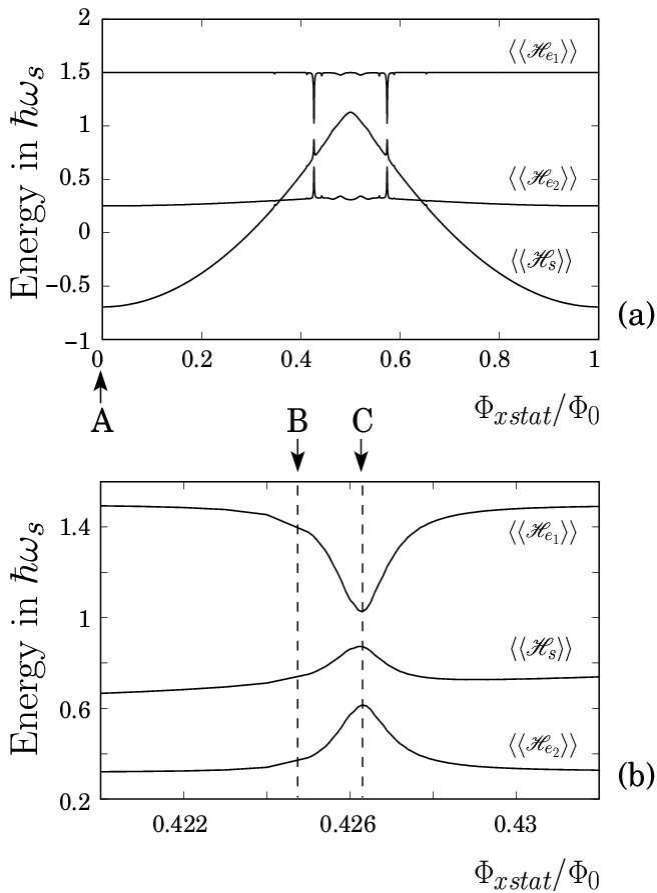}}
  \end{center} 
  \caption{(a) Time averaged energy expectation values (in units of
$\hbar\omega_{s}$) over the range $0\leq\Phi_{xstat}/\Phi_{0}\leq1$ for the
three mode system of figure~\ref{f8} and (b) as for (a), but expanded around
$\Phi_{xstat}/\Phi_{0}=0.426$. The initial state for the system is $\left|
1\right\rangle _{\mathcal{E}_{1}}\otimes\left|  \alpha\right\rangle
_{\mathcal{S}}\otimes\left|  0\right\rangle _{\mathcal{E}_{2}}$ and the
circuit parameters are as in figure~\ref{f8}.
\label{f9}
}
\end{figure}

Given a choice of the initial state for the fields in the two oscillator
modes, we can calculate the time averaged energy expectation values
(normalized in units of $\hbar\omega_{s}=\hbar/\sqrt{C_{s}\Lambda_{s}}$) of
$\mathcal{H}_{{\ }e_{1}}$, $\mathcal{H}_{e_{2}}$ and $\mathcal{H}_{s}$ using
equation (\ref{eq:average}) and integrating numerically from $0$ up to
$2\times10^{4}/\omega_{s}$. As an example, we show in figure~\ref{f9}(a) 
these time
averaged energies plotted as a function of $\Phi_{xstat}/\Phi_{0}$ assuming
that at $t=0$ the system is in the state $\left|  1\right\rangle
_{\mathcal{E}_{1}}\otimes\left|  \alpha\right\rangle _{\mathcal{S}}
\otimes\left|  0\right\rangle _{\mathcal{E}_{2}}$, i.e. with one photon in the
first em mode and none in the second. The peaks are a manifestation of the
strong coupling between the various oscillators at and around specific values
of $\Phi_{xstat}/\Phi_{0}$ where the energy transfer between the various
components of the three mode system occurs. Thus, starting in the initial
state $\left|  1\right\rangle _{\mathcal{E}_{1}}\otimes\left|  \alpha
\right\rangle _{\mathcal{S}}\otimes\left|  0\right\rangle _{\mathcal{E}_{2}}$,
it can be seen that in the exchange regions, on average, energy is being
transferred from the first mode to the SQUID ring and to the second mode. This
will become more transparent when we compute the time evolution of the
expectation values of the number operators for the components of the system
(below). In figure~\ref{f9}(b) we show one of the three mode exchange regions (around
$\Phi_{xstat}/\Phi_{0}=0.426$) of figure~\ref{f9}(a) on a much expanded scale so that
the details can be seen more clearly.

\subsection{Quantum statistics of the SQUID ring-field system}

\label{IVB}

An important aspect of non-classical electromagnetic fields is the quantum
statistics of photons (bunching of photons) described by the second order
correlations~\cite{ScullyZ97,Bachor1998}
\begin{equation}
g_{i}^{(2)}=\frac{\left\langle N_{i}^{2}\right\rangle -\left\langle
N_{i}\right\rangle }{\left\langle N_{i}\right\rangle ^{2}};\left\langle
N_{i}^{M}\right\rangle =\mathrm{Tr}[\rho_{i}(a_{i}^{\dagger}a_{i}
)^{M}];i={\ e_{1}},{e_{2}},s \label{eq:g2}
\end{equation}
The value of $g^{\left( 2\right) }=1$ corresponds to Poissonian
statistics.  Values of $g^{\left( 2\right) }$ greater than~1 indicate
photon bunching (i.e. where the photons arrive in groups) while values
of $g^{\left( 2\right) }$ smaller than one indicate antibunching (i.e.
the regular arrival of photons). The latter regime is characteristic
of non-classical electromagnetic fields since it can be shown that in
classical optics $g^{\left( 2\right) }\geq1$. Only in quantum
mechanical systems can $g^{\left( 2\right) } <
1$~\cite{ScullyZ97,Bachor1998}. This is well known in quantum optics
but is, perhaps, less familiar in condensed matter physics.

In this work we show that the statistics of photons threading the
SQUID ring affects the statistics of electron pair condensate
tunnelling through the Josephson junction in the ring. This is
quantified by the second order correlations although higher order
correlations can also be calculated~\cite{ScullyZ97}. These
correlations fully describe the quantum statistics and quantum noise
of the photons in the two field modes and the superconducting
condensate.

\subsection{Quantum Entanglement in the Three Mode System}

\label{IVC}

The creation of entangled states of multi-particle systems is a key
feature of all quantum technologies. In their pursuit the generation
of entanglements in real physical systems is clearly of very
considerable interest. In this regard, it appears that the non-linear
properties of the SQUID ring can be used very efficiently to entangle
circuit subsystems (here, field oscillator modes) that are coupled to
it. As we shall also show, the ring non-linearity can also be used
with facility to generate energy conversion between the two oscillator
field modes. Again, taking as our example the three mode system of
figure~\ref{f2}, we shall demonstrate that as this system evolves in time its
three components become, to a greater or lesser extent, entangled. The
degree of this entanglement can be quantified by using entropic
quantities. The entanglement for a two mode (ring-oscillator) system
can be quantified
by~\cite{lindbland1973,lieb_bull1975,wehrl1978,BarnettP91}
\begin{equation}
I_{AB}=S\left(  \rho_{A}\right)  +S\left(  \rho_{B}\right)  -S\left(
\rho\right)  \label{eq:e1}
\end{equation}
where $S\left(  \rho\right)  $ is the von Neumann entropy given by
\begin{equation}
S\left(  \rho\right)  =-\mathrm{Tr}\left[  \rho\ln\left(  \rho\right)
\right]
\end{equation}
with $\rho_{A}=\mathrm{Tr}_{B}\rho$ and $\rho_{B}=\mathrm{Tr}_{A}\rho$. This
entanglement entropy is positive or zero (subadditivity property of the
entropy). Examples of the calculation of this entanglement for the two mode
system can be found in our previous work~\cite{EverittSVVRPP2001}.

An analogous quantity can be used to characterize the entanglement of a three
component system~\cite{robinson67,Lanford68,Lieb73a,Lieb73b}. Thus,
for the field mode-SQUID ring-field mode system, this takes the form
\begin{equation}
I=S\left(  \rho_{{e_{1}}}\right)  +S\left(  \rho_{s}\right)  +S\left(
\rho_{{\ e_{2}}}\right)  -S\left(  \rho\right)  \label{eq:Ientrop1}
\end{equation}
which can be written as
\begin{equation}
I=I_{{e_{1}}s}+I_{s{e_{2}}}+I\left(  {e_{1}}s;s{e_{2}}\right)
\label{eq:Ientrop2}
\end{equation}
where $I_{{e_{1}}s}$ and $I_{s{e_{2}}}$ are the entanglement entropies between
$\mathcal{E}_{1}\otimes\mathcal{S}$ and $\mathcal{S}\otimes\mathcal{E}_{2}$,
as defined above (\ref{eq:e1}), and
\begin{equation}
I({e_{1}}s;s{e_{2}})=S\left(  \rho_{{e_{1}}s}\right)  +S\left(  \rho_{s{e_{2}
}}\right)  -S\left(  \rho_{s}\right)  -S\left(  \rho\right)
\label{eqIentrop3}
\end{equation}
describes a deeper entanglement between $\mathcal{E}_{1}\otimes\mathcal{S}$
and $\mathcal{S}\otimes\mathcal{E}_{2}$. Understanding of this deeper
entanglement, that exists in three component systems, is intimately connected
to the strong subadditivity property of the entropy. This can be used to
demonstrate that the quantity $I(e_{1}s;se_{2})$ is positive or zero. We note
that to prove this presented a very difficult problem in the theory
of entropy (it was a conjecture for many years until a proof was
provided~\cite{robinson67,Lanford68,Lieb73a,Lieb73b}).

In this paper we are not going to proceed further into this deep problem of
entanglement in three component systems as it has been discussed in detail
elsewhere~\cite{Vourdas92}. However, we note that the entanglement
$I$ can also be expressed as:
\begin{eqnarray}
I  &  =S\left(  \rho_{{e_{1}}}\right)  +S\left(  \rho_{s}\right)  +S\left(
\rho_{{e_{2}}}\right)  -S\left(  \rho\right) \nonumber\\
&  =I_{{e_{1}}s}+I_{{e_{1}}{e_{2}}}+I\left(  {e_{1}}s;{e_{1}}e_{2}\right)
\nonumber\\
&  =I_{{e_{2}}s}+I_{{e_{1}}{e_{2}}}+I\left(  e_{2}s;{e_{1}}{e_{2}}\right)
\end{eqnarray}
where $I\left(  {e_{1}}s;{e_{1}}{e_{2}}\right)  $ and $I\left(  {e_{2}
}s;{e_{1}}{e_{2}}\right)  $ are non-negative numbers. On physical grounds,
i.e. because we are using the SQUID ring as the intermediary between the two
field modes, we choose to show numerical results for the entanglement between
the SQUID and the first mode ($I_{{e_{1}}s}$), the SQUID and the second mode
($I_{{e_{2}}s}$) and also the $I({e_{1}}s;s{e_{2}})$.

\section{Numerical Calculations}

As we have shown in figures~\ref{f6} and~\ref{f9}, strong coupling between the various of
components of ring-field mode systems only occurs over small regions in
$\Phi_{xstat}$ - the exchange regions. We now see how the variation in
coupling across an exchange region affects the number operator expectation
values, quantum statistics and entanglements - all important quantities
reflecting on the quantum behaviour of these systems. Continuing from figure~\ref{f9}, 
we calculate these quantities at each of the three flux bias points A, B
and C (at $\Phi_{xstat}/\Phi_{0}=0.0,0.4246$ and $0.4263$, respectively). In
each of the following computed examples we assume that at $t=0$ the first
field mode $\mathcal{E}_{1}$ contains one or more photons while the second
contains none. In our first set of examples we choose the $t=0$ state in the
first mode to be a number state; in the second set we make this a coherent
state $|A\rangle_{\mathcal{E}{_{1}}}$, where $a_{{e_{1}}}|A\rangle
_{\mathcal{E}{_{1}}}=A|A\rangle_{\mathcal{E}{_{1}}}$. For the case of the
number state we assume that at $t=0$ the three mode system is in the state
$|1\rangle_{\mathcal{E}{_{1}}}\otimes|\alpha\rangle_{\mathcal{S}}
\otimes|0\rangle_{\mathcal{E}{_{2}}}$. For the example where we adopt a
coherent state for the first mode we choose for illustrative purposes (and
computational ease) the system state $|A=i\sqrt{3}\rangle_{\mathcal{E}{_{1}}
}\otimes|\alpha\rangle_{\mathcal{S}}\otimes|0\rangle_{\mathcal{E}{_{2}}}$.

\subsection{Number state computations}

\begin{figure}[tb!]
  \begin{center}
    \resizebox*{0.48\textwidth}{!}{\includegraphics{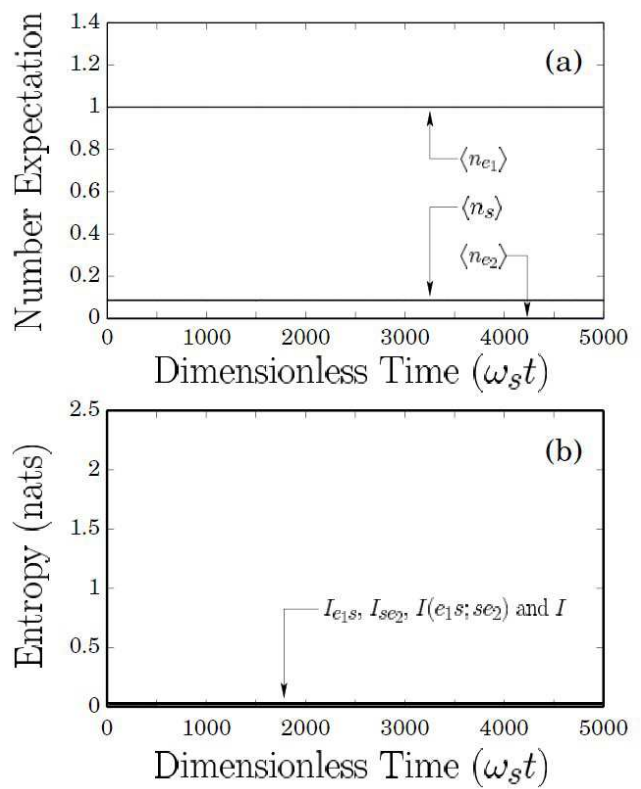}}
  \end{center} 
  \caption{ Starting in a pure state $|1\rangle_{\mathcal{E}_{1}}
\otimes|\alpha\rangle_{\mathcal{S}}\otimes|0\rangle_{\mathcal{E}_{2}}$, plots
of (a) the number expectation values $\left\langle n_{e_{1}}\right\rangle $,
$\left\langle n_{s}\right\rangle $, $\left\langle n_{e_{2}}\right\rangle $ and
(b) the entanglement entropies $I_{e_{1}s}$, $I_{se_{2}}$, $I\left(
e_{1}s;se_{2}\right)  $, $I$ versus dimensionless time $\omega_{s}t$ for the
three component system of figure~\ref{f9} with the ring flux biased at A in figure~\ref{f9}(a). 
Here, the system parameter values are as for figure~\ref{f8}(b).
\label{f10}
}
\end{figure}

\begin{figure}[tb!]
  \begin{center}
    \resizebox*{0.48\textwidth}{!}{\includegraphics{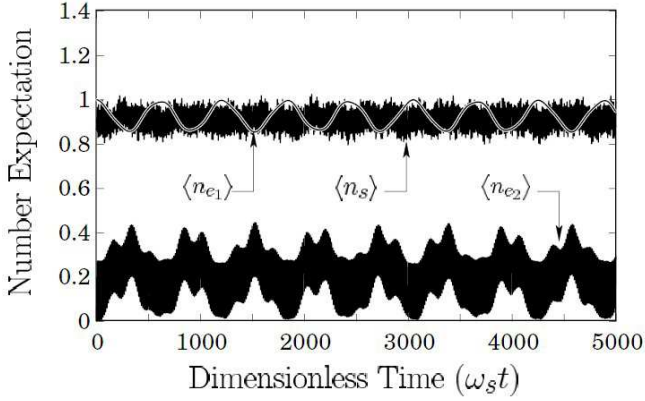}}
  \end{center} 
  \caption{ With the initial pure state $|1\rangle_{\mathcal{E}_{1}
}\otimes|\alpha\rangle_{\mathcal{S}}\otimes|0\rangle_{\mathcal{E}_{2}}$, and
with the parameter values of figure~\ref{f8}(b), a plot of the number expectation
values $\left\langle n_{e_{1}}\right\rangle $, $\left\langle n_{s}
\right\rangle $, $\left\langle n_{e_{2}}\right\rangle $ versus dimensionless
time $\omega_{s}t$ for the static magnetic flux on the SQUID ring set at point
B in figure~\ref{f9}(b).
\label{f11}
}
\end{figure}

\begin{figure}[tb!]
  \begin{center}
    \resizebox*{0.48\textwidth}{!}{\includegraphics{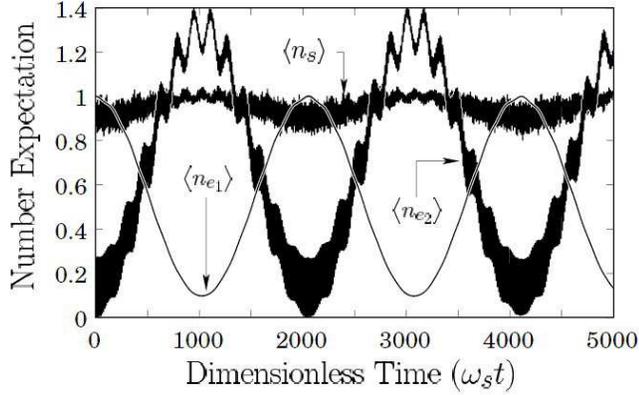}}
  \end{center} 
  \caption{ Starting in the pure state $|1\rangle_{\mathcal{E}_{1}
}\otimes|\alpha\rangle_{\mathcal{S}}\otimes|0\rangle_{\mathcal{E}_{2}}$, and
with the parameter values of figure~\ref{f8}(b), plots of the number expectation
values $\left\langle n_{e_{1}}\right\rangle $, $\left\langle n_{s}
\right\rangle $, $\left\langle n_{e_{2}}\right\rangle $ versus dimensionless
time $\omega_{s}t$ for the three component system of figure~\ref{f9} with the SQUID
ring flux biased at point C in figure~\ref{f9}(b).
\label{f12}
}
\end{figure}

As is evident from figures~\ref{f9}(a) and~\ref{f9}(b), the flux bias points have been
selected either to be well away from, or within, an exchange region, i.e.
point A and points B and C, respectively. For a complete, quantitative view of
the system we should compute, in sequence, the number expectation values
$\left\langle n_{{e_{1}},s,{e_{2}}}\right\rangle $, the entropies $\left(
I_{e_{1}s},I_{se_{2}},I\left(  {e_{1}}s;s{e_{2}}\right)  \mathrm{\ and}\text{
}I\right)  $and the $g_{{e_{1}},s,{e_{2}}}^{\left(  2\right)  }$ correlations
for the first field mode $\left(  \mathcal{E}_{1}\right)  $, the SQUID ring
$\left(  \mathcal{S}\right)  $ and the second field mode $\left(
\mathcal{E}_{2}\right)  $ as a function of normalized time $\omega_{s}t$.
However, it is apparent in figure~\ref{f10}
 (bias point A) that, starting in a pure
state $|1\rangle_{\mathcal{E}{_{1}}}\otimes|\alpha\rangle_{\mathcal{S}}
\otimes|0\rangle_{\mathcal{E}{_{2}}}$ for $\mathcal{E}_{1}$, the number
expectation values (a) and entropies (b) remain constant as a function of
time. We note that $\left\langle n_{s}\right\rangle $ is not zero because the
ground state $\left|  \alpha\right\rangle _{\mathcal{S}}$ of the SQUID ring is
not the same as the ground state of a simple harmonic oscillator $\left|
0\right\rangle $. From the definition given in section \ref{IVB}, the fact
that $\left\langle n_{e_{2}}\right\rangle =0$ makes the calculation of
$g^{\left(  2\right)  }$ for the second mode at bias point A physically
unmeaningful since this is division by zero. It is therefore very sensitive to
numerical error. However, such is not the case when the bias point point is
shifted into an exchange region. Starting again with the system state
$|1\rangle_{\mathcal{E}{_{1}}}\otimes|\alpha\rangle_{\mathcal{S}}
\otimes|0\rangle_{\mathcal{E}{_{2}}}$, we show in figures~\ref{f11} 
(bias point B)
and~\ref{f12} (bias point C) the average number of quanta - the $\left\langle
n_{i}\right\rangle $ - in each of the three modes as a function of time. With
this choice of starting state these results demonstrate clearly the
quasi-periodic exchange of energy between the various components of the
system. Since the exchange coupling is strongest at C, this is where we would
expect to find the maximum energy transfer between the first and second field
modes, as is the case (figure~\ref{f12}).
 We note that in the computed results of
figure~\ref{f12} the second field mode number expectation value (and that of the
SQUID ring to a much smaller extent) is a maximum when that for the first
field mode is a minimum. This is the signature for frequency (down)
conversion, in this example by a factor of two. The process could, of course,
be run backwards to generate frequency up conversion from the second to the
first field mode via the quantum non-linearity of the SQUID ring. Given that
this non-linearity can be to all orders, we see no obvious reason why much
higher ratio frequency conversions should not prove practicable.

From a theoretical viewpoint the problem with demonstrating high ratio
frequency conversions is the rapid rise in the number of basis states
required as the down (up) conversion frequency ratio increases. The
computational difficulties increase accordingly. Nevertheless, even
given the limitations on the computational power we have available
(Compaq XP1000 alphaserver with 2GB RAM), we have been able to
demonstrate quantum down conversion by a factor of ten in frequency.
We intend to deal with this in a future publication.  There is some
indication that these down conversion processes
occur~\cite{WhitemanCPPSRD98}.  In figure~\ref{f12} the input state is
the number state $|1\rangle_{\mathcal{E}_{1}}$ but, as we shall show,
down conversion can occur for a coherent input state. It may well be
that this ability to generate photon down/up conversion could have
practical application for pure state sources in quantum information
processing and quantum computing. For example, it may prove desirable
to take single photon terahertz sources, as are now being
developed~\cite{FaistCSSHC94,amone2000}, and use these to provide the
input state to a SQUID ring to generate photons at much lower
frequencies suitable for solid state quantum circuit technologies. It
is also clear that if very large down/up frequency conversion ratios
can be achieved experimentally, there could well be interesting
metrological applications, for example, in frequency standards.

\begin{figure}[tb!]
  \begin{center}
    \resizebox*{0.48\textwidth}{!}{\includegraphics{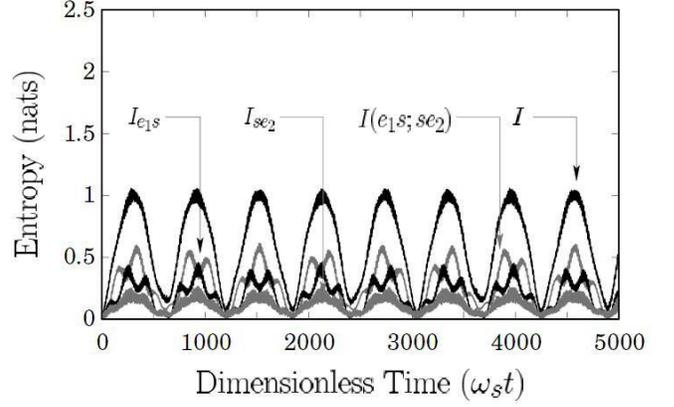}}
  \end{center} 
  \caption{ Entanglement entropies $I_{{e_{1}}s}$, $I_{s{e_{2}}}$,
$I\left(  e_{1}s;se_{2}\right)  $ and $I$ versus dimensionless time
$\omega_{s}t$ for the three component system of figure~\ref{f9}, with the parameter
values of figure~\ref{f8}(b), starting in state $|1\rangle_{\mathcal{E}_{1}}
\otimes|\alpha\rangle_{\mathcal{S}}\otimes|0\rangle_{\mathcal{E}_{2}}$ with
the SQUID ring flux set at point B [figure~\ref{f9}(b)].
\label{f13}
}
\end{figure}

\begin{figure}[tb!]
  \begin{center}
    \resizebox*{0.48\textwidth}{!}{\includegraphics{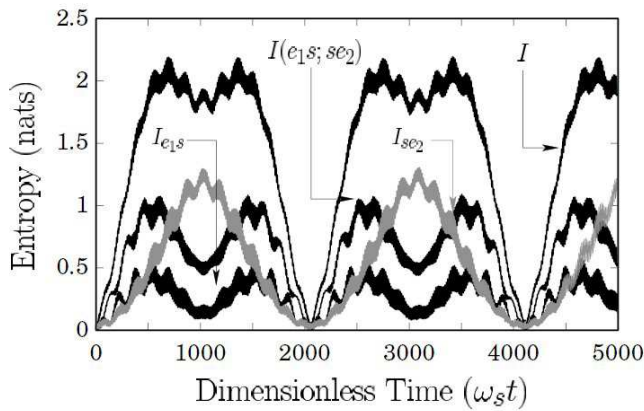}}
  \end{center} 
  \caption{Entanglement entropies $I_{{e_{1}}s}$, $I_{s{e_{2}}}$,
$I\left(  e_{1}s;se_{2}\right)  $ and $I$ versus dimensionless time
$\omega_{s}t$ for the three component system of figure~\ref{f9}, with the parameter
values of figure~\ref{f8}(b), starting in state $|1\rangle_{\mathcal{E}_{1}}
\otimes|\alpha\rangle_{\mathcal{S}}\otimes|0\rangle_{\mathcal{E}_{2}}$ with
the SQUID ring flux biased at point C in figure~\ref{f9}(b).
\label{f14}
}
\end{figure}

This role in linking the two field modes together in a strongly
non-linear, quantum mechanical manner is emphasized in figures~\ref{f13} 
and~\ref{f14}
. 
Here, the time varying entanglement entropies are computed for
bias points B (figure~\ref{f13}) and C (figure~\ref{f14}) following the definitions
given in section \ref{IVC}. Here, again, we have started the system in
state $|1\rangle_{\mathcal{E}{_{1}}
  }\otimes|\alpha\rangle_{\mathcal{S}}\otimes|0\rangle_{\mathcal{E}{_{2}}}$.
It can be seen that the entanglement between the various components of
the system ($I_{{e_{1}}s}$, $I_{s{e_{2}}}$ and $I\left(
  {e_{1}}s;s{e_{2}}\right) $), and the total entanglement entropy for
the system $\left( I\right) $, are stronger at C than at B which, from
figures~\ref{f9}(b),~\ref{f11} and~\ref{f12}, is to be expected. In our opinion it is this
ability to control the degree of entanglement between the components
of (for example) this three mode system simply by changing
$\Phi_{xstat}$ which marks out the SQUID ring as a potentially very
useful device in future quantum circuit technologies. This is
emphasized by the contrast between figures~\ref{f10}(b) (for bias point A)
and~\ref{f14}, where the system and subsystem entanglement go from zero to a
maximum for an adjustment in $\Phi_{xstat}$ around $0.04\Phi_{0}$.

Although, in the above we have considered in some detail entanglement
between two field modes (input and output) interacting via a SQUID
ring, there are many other coupled systems of field modes and SQUID
rings which could be studied. One which may be of importance, both
scientifically and technologically, is an input field mode linked
through a SQUID ring to two separate output modes at half the input
frequency. From the results obtained in this paper we would expect the
two (down converted) photons to be strongly entangled with the degree
of entanglement controlled, again, by the bias flux $\Phi_{xstat}$
applied to the ring. Furthermore, given the non-linearity of the SQUID
ring, we would also expect it to be possible to entangle a large
number of output photons starting at an initial input frequency and
down converting to a whole set of lower frequency output modes.  As a
technique, the use of a SQUID ring to generate entanglements between
several systems could well be applied to great advantage in
fundamental experimental studies of quantum
mechanics~\cite{Greenberger1990,Mermin1990,Macchiavello1999}. It could
also have implications in quantum computing, for example, in creating
an entangled input register for a quantum computer. It has also been
suggested that the creation of (large number) multi-particle entangled
systems could lead to new sensors and instrumentation of unparalleled
sensitivity~\cite{Brooks1999} and it may be that SQUID rings are very
well suited to creating these entanglements, at least for photons.

There are other possible ways that the input register of a quantum
computer could be based on the non-linear properties of SQUID rings
described in this paper. For example, we could set the three modes of
the coupled system in figure~\ref{f2} all to have the same oscillator
frequency $\omega_{s}$.  Then, with $\Phi_{xstat}$ biased within an
exchange region, we could arrange to create a qubit superposition
state of $\left| 0\right\rangle $ and $\left| 1\right\rangle $ in the
output mode starting from the number state $\left| 1\right\rangle $ of
the input mode. As our results have demonstrated, this could be done
in such a way as to ensure that the input and output oscillator modes
are entangled.  Once the desired qubit state of the output mode had
been realised, in principle the bias flux could then be switched away
rapidly from the exchange region (or switched off) thus leaving the
input and output modes entangled but uncoupled. An array of these
circuits could then be used as an qubit register for a quantum
computer, where the qubits would be entangled but not coupled to the
input modes. Conceivably, this arrangement could facilitate quantum
error correction for quantum
computation~\cite{Williamsclearwater1998}.  Schemes of this kind may
well find application in quantum encryption and transmission of
information at a more complex level than is usually
considered~\cite{Lopopescuspiller,Williamsclearwater1998,OpcitBrooks1999}.

\begin{figure}[tb!]
  \begin{center}
    \resizebox*{0.48\textwidth}{!}{\includegraphics{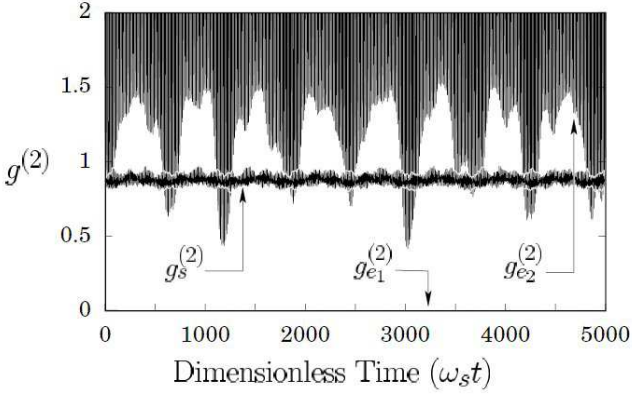}}
  \end{center} 
  \caption{Second order correlations $g_{{e_{1}}}^{\left(  2\right)
},g_{s}^{\left(  2\right)  }$ and $g_{{e_{2}}}^{\left(  2\right)  }$ versus
dimensionless $\omega_{s}t$ for the three component system of figure~\ref{f9},
starting in a pure state $|1\rangle_{\mathcal{E}_{1}}\otimes|\alpha
\rangle_{\mathcal{S}}\otimes|0\rangle_{\mathcal{E}_{2}}$ with the static
magnetic flux on the SQUID ring set at bias point B in figure~\ref{f9}(b). Here, the
system parameter values are as for figure~\ref{f8}(b).
\label{f15}
}
\end{figure}

\begin{figure}[tb!]
  \begin{center}
    \resizebox*{0.48\textwidth}{!}{\includegraphics{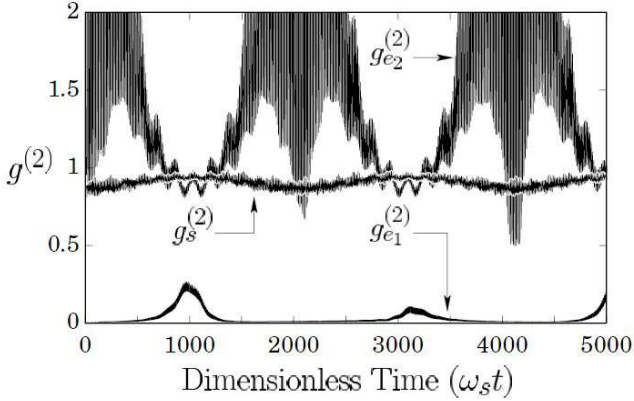}}
  \end{center} 
  \caption{ Second order correlations $g_{{e_{1}}}^{\left(  2\right)
},g_{s}^{\left(  2\right)  }$ and $g_{{e_{2}}}^{\left(  2\right)  }$ versus
dimensionless $\omega_{s}t$ for the three component system of figure~\ref{f9}
[parameter values as for figure~\ref{f8}(b)], starting in a pure state $|1\rangle
_{\mathcal{E}_{1}}\otimes|\alpha\rangle_{\mathcal{S}}\otimes|0\rangle
_{\mathcal{E}_{2}}$ with the SQUID ring bias flux set at bias point C in
figure~\ref{f9}(b).
\label{f16}
}
\end{figure}

Since the underlying purpose of this work is to demonstrate the influence of
the SQUID ring non-linearity on a coupled quantum system, it is important to
show quantitatively the way in which the quantum statistics of the photons
affects the quantum statistics of the electron pairs (i.e. the superconducting
condensate flowing through the weak link in the ring). As we explained above
(section \ref{IVB}), this is quantified with the second order correlations
$g_{i}^{\left(  2\right)  }$. In figures~\ref{f15} 
and~\ref{f16} 
we plot the second order
correlations $g_{{e_{1}}}^{\left(  2\right)  }$, $g_{s}^{\left(  2\right)  }$
and $g_{{e_{2}}}^{\left(  2\right)  }$ for bias points B and C as functions of
time with, again, a starting state for the system of $|1\rangle_{\mathcal{E}
{_{1}}}\otimes|\alpha\rangle_{\mathcal{S }}\otimes|0\rangle_{\mathcal{E}{_{2}
}}$. As with the number expectation values and the entanglement entropies, we
see a strong oscillatory behaviour, particularly in figure~\ref{f16} (biased at point
C). In order to interpret the results we first note that for the state
$|0\rangle_{i}$ $\left(  i=\mathcal{E}{_{1}},\mathcal{E}{_{2}},\mathcal{S}
\right)  $, or other states close to this state, the average number of photons
is near zero. From equation (\ref{eq:g2}) this means that the second order
correlation becomes very large. We also note for the number state $|1\rangle$,
$\left\langle N^{2}\right\rangle -\left\langle N\right\rangle =0$ and the
corresponding $g^{\left(  2\right)  }=0$. With this in mind, we see in figure~\ref{f15} 
(bias point B) that the first field mode, in number state $|1\rangle
_{\mathcal{E}{_{1}}} $ at $t=0$, starts with a $g^{\left(  2\right)  }$ at
zero. It remains extremely close to this value over the time of the
computation, i.e. this field mode stays reasonably close to the number state
$|1\rangle_{\mathcal{E }{_{1}}}$. By contrast, the second field mode, assumed
to be in the state $|0\rangle_{\mathcal{E}{_{2}}}$ at $t=0$, has on average a
large $g^{\left(  2\right)  }$ value, although this dips well below unity
almost periodically with time. This demonstrates that even at bias point B, on
the edge of the exchange region, the value of $g_{e_{2}}^{\left(  2\right)  }$
regularly falls below one. It also shows that for this initial condition the
quantum statistics of the second field mode cannot be described by classical
means. In figure~\ref{f16} (bias point C), at $t=0$, we again assume that the first
field mode is in state $|1\rangle_{\mathcal{E}{_{1}}}$ with the second in
state $|0\rangle_{\mathcal{E}{_{2}}}$. As before, the first field mode starts
at $g_{{e_{1}}}^{\left(  2\right)  }=0$ but as the wavefunction for the system
evolves with time we see that $g_{{e_{1}}}^{\left(  2\right)  }$ regularly
shifts away from zero. Correspondingly, the first field mode is no longer in
the pure number state $|1\rangle_{\mathcal{E}_{1}}$ due to its interaction
with the rest of the system$.$ The first field mode is described by a reduced
density operator which, at these points, represents statistical mixture of
states with a low photon number expectation value (as can be seen from figure~\ref{f12}). 
As a consequence, $g_{{e_{1}}}^{\left(  2\right)  }$ increases since the
denominator in equation (\ref{eq:g2}) becomes very small around these points.

\subsection{Coherent state computations}

\begin{figure}[tb!]
  \begin{center}
    \resizebox*{0.48\textwidth}{!}{\includegraphics{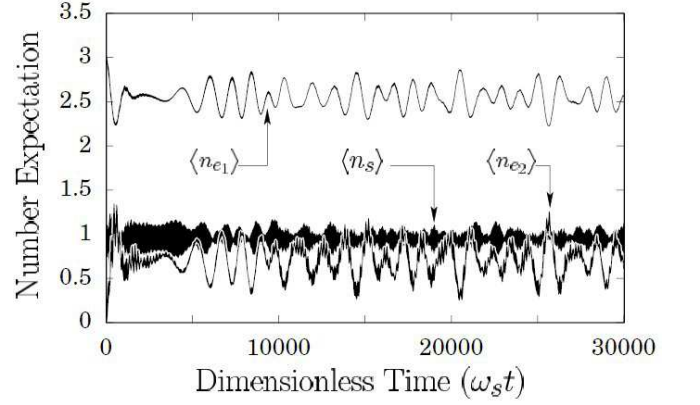}}
  \end{center} 
  \caption{Starting the three component system with the first field mode
in a coherent state, i.e. $|A=i\sqrt{3}\rangle_{\mathcal{E}_{1}}\otimes
|\alpha\rangle_{\mathcal{S}}\otimes|0\rangle_{\mathcal{E}_{2}}$, and using the
parameter values of figure~\ref{f8}(b), plots of the number expectation values
$\left\langle n_{e_{1}}\right\rangle $, $\left\langle n_{s}\right\rangle $,
$\left\langle n_{e_{2}}\right\rangle $ against dimensionless time $\omega
_{s}t$ for the static flux on the SQUID ring set at point C in figure~\ref{f9}(b).
\label{f17}
}
\end{figure}

\begin{figure}[tb!]
  \begin{center}
    \resizebox*{0.48\textwidth}{!}{\includegraphics{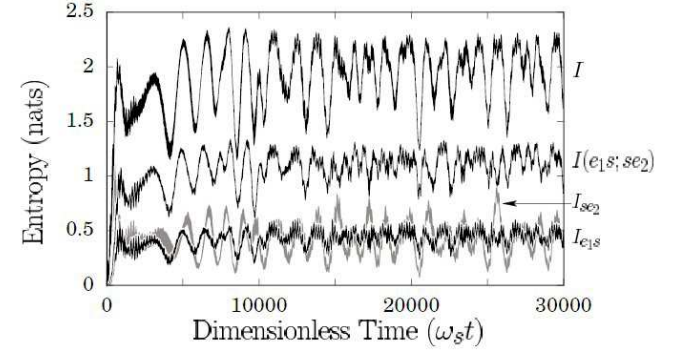}}
  \end{center} 
  \caption{With initial state $|A=i\sqrt{3}\rangle_{\mathcal{E}_{1}
}\otimes|\alpha\rangle_{\mathcal{S}}\otimes|0\rangle_{\mathcal{E}_{2}}$ for
the three component system, the entanglement entropies $I_{{e_{1}}s}$,
$I_{s{\ e_{2}}}$, $I\left(  e_{1}s;se_{2}\right)  $ and $I$ versus
dimensionless time $\omega_{s}t$, where the system parameters are as for
figure~\ref{f8}(b) and the SQUID ring is flux biased at point C in figure~\ref{f9}(b).
\label{f18}
}
\end{figure}

\begin{figure}[tb!]
  \begin{center}
    \resizebox*{0.48\textwidth}{!}{\includegraphics{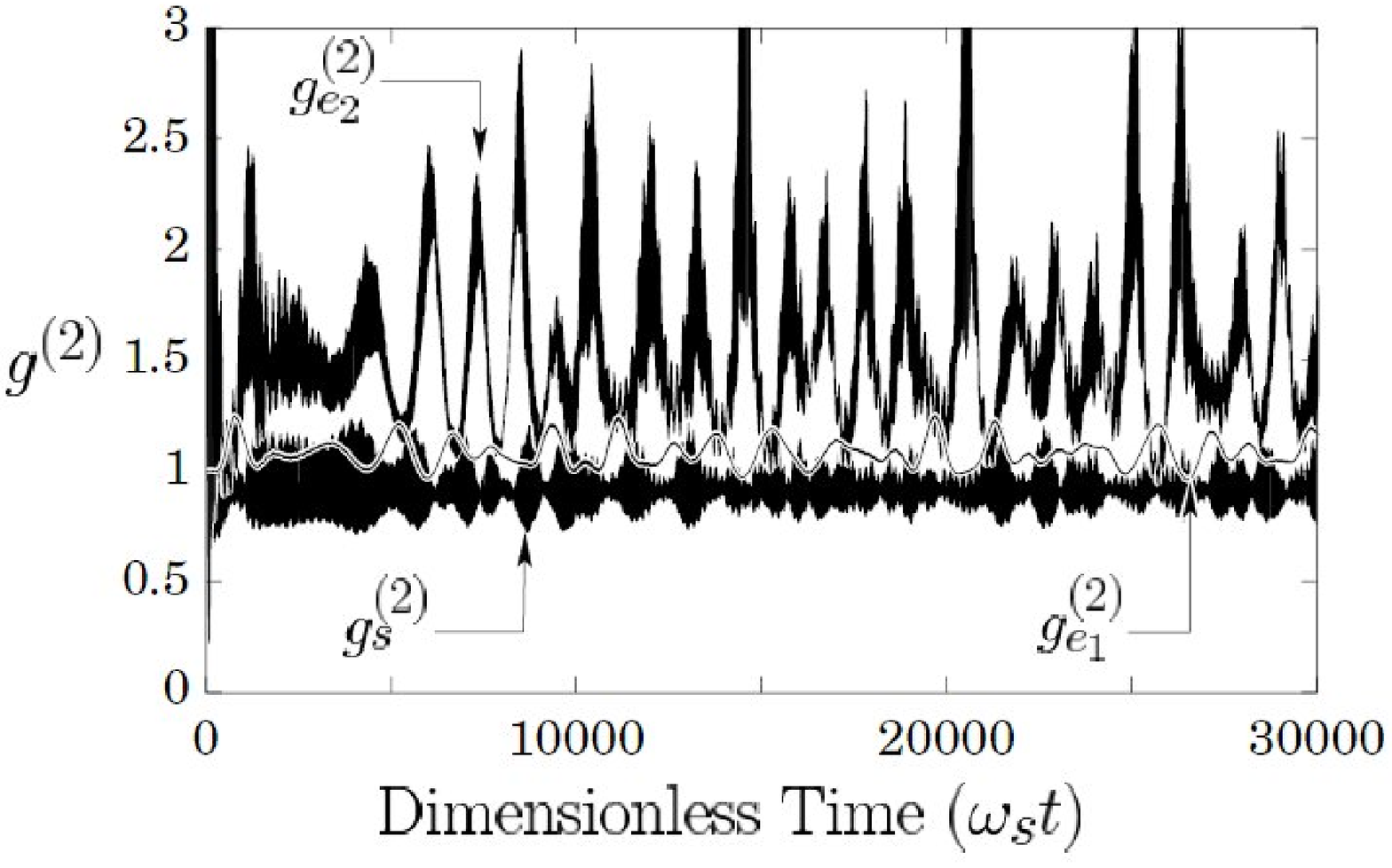}}
  \end{center} 
  \caption{Second order correlations $g_{{e_{1}}}^{\left(  2\right)
},g_{s}^{\left(  2\right)  }$ and $g_{{e_{2}}}^{\left(  2\right)  }$ versus
dimensionless time $\omega_{s}t$ for the three component system [parameter
values as for figure~\ref{f8}(b)] starting in state $|A=i\sqrt{3}\rangle
_{\mathcal{\ E}_{1}}\otimes|\alpha\rangle_{\mathcal{S}}\otimes|0\rangle
_{\mathcal{E} _{2}}$ with the flux on the SQUID ring set at bias point C in
figure~\ref{f9}(b).
\label{f19}
}
\end{figure}

In figures~\ref{f17},
~\ref{f18} 
and~\ref{f19} 
we show the number expectation values, the
entanglement entropies and the $g^{\left( 2\right) }$ correlations for
the bias point C in figure~\ref{f9}(b), taking the initial state of the
system as
$|A=i\sqrt{3}\rangle_{\mathcal{E}{_{1}}}\otimes|\alpha\rangle_{\mathcal{\ 
    S} }\otimes|0\rangle_{\mathcal{E}{_{2}}}$, i.e. where the first
field mode is in the coherent state
$|i\sqrt{3}\rangle_{\mathcal{E}{_{1}}}$ at $t=0$ . It is apparent
(figure~\ref{f17}) that there is energy transfer, via the SQUID ring, between
the first and second field modes of the system, just as in figure~\ref{f12}
for the pure state $|1\rangle_{\mathcal{E}{_{1}}}\otimes|\alpha\rangle
_{\mathcal{S}}\otimes|0\rangle_{\mathcal{E}{_{2}}}$ . Thus, when
$\left\langle n_{{e_{1}}}\right\rangle $ decreases $\left\langle
  n_{{e_{2}}}\right\rangle $ increases, and vice versa. However, it is
evident that the regular oscillatory behaviour seen in figures~\ref{f11}
and~\ref{f12} has been lost. We note that, as in figures~\ref{f11} and~\ref{f12}, the
number expectation value of the SQUID ring remains at a roughly
constant value, highlighting the view that the SQUID ring is acting as
a non-linear control medium linking the two quantum field modes
together. In figure~\ref{f18} we see that the components of the system again
entangled very strongly but, unlike the previous computations of
figures~\ref{f13} and~\ref{f14}, there is no longer any quasi-periodic
disentanglement to be seen. The system remains entangled at all points
in time, i.e. the total entanglement entropy is always high. In
figure~\ref{f19} there is clearly a significant deviation from the behaviour
of the $g^{\left( 2\right) }$ coefficients displayed in figures~\ref{f15} and~\ref{f16}.  
Thus, in figure~\ref{f19}, $g_{{e_{1}}}^{\left( 2\right) }$ is close to
one for all of the time evolution (and for most of the time just
greater than one) whereas $g_{s}^{\left( 2\right) }$ spends most of
its time just less than one and $g_{{e_{2}}}^{\left( 2\right) }$ is
almost always greater than one.

In all the examples given above we observe that, as expected, the system
displays strong entanglement when the first field mode has a low number
expectation value, i.e. the components of the system have evolved away from
their initial pure states. We also note that when we start the first field
mode in the pure state $|1\rangle_{\mathcal{E}{_{1}}}\otimes|\alpha
\rangle_{\mathcal{S}}\otimes|0\rangle_{\mathcal{E}{_{2}}}$, at bias points B
and C, we find strong entanglement between the various components of the
coupled system which, at certain times (semi-periodically), disentangle. The
entanglement entropies are, of course, theoretical quantities that demonstrate
the development of quantum correlations between the three modes. In principle,
experimental observation of these quantum correlations can be achieved by
determining Bell type
inequalities~\cite{ScullyZ97,Bell1966,Williamsclearwater1998}. In the
context of the work presented here this will require further theoretical
investigation. In figures~\ref{f15},~\ref{f16} and~\ref{f19} it is evident that the $g^{\left(
2\right)  }$ correlation coefficient for at least one of the components of the
system becomes less than one at some point in the evolution of the system.
Hence, for the initial conditions used in this work, we conclude that the
photon statistics of this system cannot be described by classical optics.

\section{Conclusions}

\label{VI}

In this paper we have studied the coupling of a SQUID ring to two em field
modes. For this we have made the assumption that the ambient temperature of
the system is low enough to be able to treat each of the three components
(ring + two field modes) quantum mechanically. Our purpose has been to
demonstrate that the SQUID ring, as a non-linear quantum object, can be used
to couple a number of quantum oscillators together to generate physical
phenomena of great interest. As we have emphasized, there exist obvious
parallels with the field of quantum optics. However, in quantum optical
systems the coupling media involved generally display only weak polynomial
non-linearity. In contrast the SQUID ring, with the cosine term in its
Hamiltonian description, can generate non-linear interactions to all orders.
It is this, plus the $\Phi_{0}$-periodic nature of its behaviour as a function
of external flux $\Phi_{x}$, which makes the SQUID ring of such interest in
the burgeoning field of quantum circuit technology. Viewed from the
perspective of quantum optics, the SQUID ring (or a set of coupled SQUID
rings) can be thought of as a non-linear medium par excellence which can
easily create very strongly coupled regimes (albeit at lower frequencies - for
example, at THz frequencies and below) which are inaccessible using
conventional optical materials.

In our theoretical investigations of the two em field modes coupled through a
quantum SQUID ring we have also applied a static external magnetic flux
$\left(  \Phi_{xstat}\right)  $ to the ring. In this paper we started with the
simpler example of a SQUID ring interacting with a single em mode. We showed
that the coupling between the components of the system can be strong. This
strong coupling only occurs over small ranges in $\Phi_{xstat}$, centred
around specific values of this bias flux, i.e. in what we term exchange
regions which are govern by the energy eigenstructure of the system. It is in
and around the exchange regions that the quantum non-linear nature of the
SQUID ring is made manifest through the coupling of the field modes via the
ring. For example, in these regions energy can be exchanged between the field
modes and the SQUID ring and, through the intermediary of the ring, between
the field modes themselves. From this we suggest that, suitably driven, this
system may act as a frequency converter suitable for operation up to the THz
range. From our viewpoint this illustrates the utility of these exchange
regions since it is here that the strong quantum couplings develop between the
components of the system. We have added to this perspective by calculating
other physical phenomena associated with such a coupled quantum system. To
illustrate this we have computed the statistics of the various quanta
(quantified with the second order correlations) and the degree of entanglement
(quantified with various entropies) between the components of the system, both
outside and within the exchange regions. Our results demonstrate quite clearly
that the mesoscopic SQUID ring can be used as a flux tunable element to
manipulate these (and presumably other) quantum properties of these coupled
circuit systems. As such, the work presented here may be of considerable
relevance to current experiments on quantum superposition of states in SQUID
rings and on probing crossing/anticrossing regions of their energy level
structure~\cite{FriedmanPCTL00,vanderWalWSHOLM00,Ralph2001}.

We note that we have neglected dissipation in the results presented in
this paper. Thus, we have computed the time evolution of the three
mode system using the equation $\partial_{t}\rho=-i[H,\rho]$. A more
realistic calculation, with dissipation due to the environment taken
into account (for example, in the
references~\onlinecite{LeggettCDFGZ87,weiss1999,giulini1996}) is now being
developed. The equation for the time evolution then takes the form
$\partial_{t}\rho=-i[H,\rho]+f(\rho)$, where a dissipative term
$f(\rho)$ has been introduced to represent the environment. We intend
to extend our work to investigate in detail the effects of this
environmental dissipation on the behaviour of SQUID ring-field mode
systems.

We suggest that the three component system (SQUID ring + two field modes), and
its extensions, is rich in possibilities for device applications (e.g. in
quantum gates, quantum encryption and frequency conversion); it is also a
pointer to more sophisticated quantum technologies in the
future~\cite{Lopopescuspiller}. Given that the technical problems
associated with such technologies can be overcome, it seems likely that the
SQUID ring (and related weak link circuits) will, in the future, be able to
operate at THz frequencies. This could complement the current drive to develop
THz applications in (classical) communications and
imaging~\cite{FaistCSSHC94,amone2000}. It would also allow for
quantum circuit technologies to be utilized at quite accessible temperatures.

\begin{acknowledgments}

We would like to thank to the NESTA Organisation and the EPSRC for their generous funding of
this work. We would also like to express our thanks to Professors C.H. van de
Wal and A. Sobolev for very useful discussions.
\end{acknowledgments}

\end{document}